\documentclass[structabstract]{aa} 
\usepackage{natbib}	

\usepackage{txfonts,epsfig,graphicx,natbib,url,twoopt}
\usepackage[breaklinks=true]{hyperref} 
\hypersetup{colorlinks=true,citecolor=blue}
\bibpunct{(}{)}{;}{a}{}{,}  
\usepackage[none]{hyphenat}
\usepackage[usenames,dvipsnames]{color}

\begin{document}

   \title{Trapping dust particles in the outer regions of protoplanetary disks}

   \author{Pinilla P.\inst{1,2}
          Birnstiel T. \inst{3,4}
          Ricci L.\inst{5,6}
          Dullemond C. P.\inst{1, 2}
          Uribe A. L. \inst{2}
          Testi L.\inst{6, 7}
          \and
          Natta A.\inst{7}
          }
   \institute{Zentrum f\"ur Astronomie der Universit\"at Heidelberg, Institut f\"ur Theoretische Astrophysik, Albert-Ueberle-Str. 2, 69120 Heidelberg, Germany\\
              \email{pinilla@uni-heidelberg.de}
              \and
              Max-Planck-Institut f\"ur Astronomie, K\"onigstuhl 17, 69117 Heidelberg, Germany
              \and
              University Observatory Munich, Scheinerstr. 1, D-81679 M\"unchen, Germany
              \and
              Excellence Cluster Universe, Boltzmannstr. 2, D-85748 Garching, Germany
             \and
             Department of Astronomy, California Institute of Technology, MC 249-17, Pasadena, CA 91125, USA
             \and
             European Southern Observatory, Karl-Schwarzschild-Strasse 2, 85748 Garching, Germany 
             \and
             INAF - Osservatorio Astrofisico di Arcetri, Largo Fermi 5, I-50125 Firenze, Italy
             }

   \date{Received  October 5 2011/ Accepted December 5 2011}

%%%%%%%%%%%%%
% ABSTRACT
%%%%%%%%%%%%%
 
  \abstract
  % context heading (optional)
   {Dust particles are observed at mm sizes in outer regions of the disk, although theoretically, radial drift does not allow dust particles to form pebbles.}
  % aims heading (mandatory)
   {In order to explain  grain growth to mm sized particles and their retention in outer regions of protoplanetary disks, as it is observed at sub-mm and mm wavelengths, we investigate if strong inhomogeneities in the  gas density profiles can slow down excessive radial drift and can help dust particles to grow.}
  % methods heading (mandatory)
   {We use coagulation/fragmentation and disk-structure models, to simulate the evolution of dust in a bumpy surface density profile which we mimic with a sinusoidal disturbance. For different values of the amplitude and length scale of the bumps, we investigate the ability of this model to produce and retain large particles on million years time scales. In addition,  we introduced a comparison between the pressure   inhomogeneities  considered in this work and the pressure profiles that come from magnetorotational instability. Using the Common Astronomy Software Applications ALMA simulator, we study if   there are observational signatures of these pressure inhomogeneities that can be seen with ALMA.}
  % results heading (mandatory)
   {We present the favorable conditions to trap dust particles and the corresponding calculations predicting the spectral slope in the mm-wavelength range, to compare with current observations. Finally we present simulated images using different antenna configurations of ALMA  at different frequencies, to show that  the ring structures will be detectable at the distances of the Taurus Auriga or Ophiucus star forming regions.}
  {}
  

   \authorrunning{Pinilla P. et al.}
   \maketitle

%%%%%%%%%%
\section{Introduction}     \label{sec1}
%%%%%%%%%%

The study of planet formation is an important field in astronomy with an increasing research since the middle of the twentieth century,  however there are still countless unanswered questions. One of these questions is the observed grain growth to mm sized particles in the outer disk regions (\cite{1991ApJ...381..250B};  \cite{2000ApJ...534L.101W, 2005ApJ...626L.109W}; \cite{2001ApJ...554.1087T,  2003A&A...403..323T}; \cite{2005ApJ...631.1134A}; \cite{2006A&A...446..211R}; \cite{2007prpl.conf..767N}; \cite{2009ApJ...701..260I}; \cite{2009A&A...495..869L}; \cite{2010A&A...512A..15R, 2011A&A...525A..81R}; \cite{2011A&A...529A.105G}) that suggests a mechanism operating that prevents the rapid inward drift  (\cite{1997Icar..128..213K}, \cite{2007A&A...469.1169B}, \cite{2007Natur.448.1022J}). Different efforts are aimed to explain theoretically the growth from small dust particles to planetesimals, which have led to the development of different numerical models, e.g. \cite{1981Icar...45..517N},  \cite{2005A&A...434..971D}, \cite{2008A&A...480..859B}, \cite{2008A&A...489..931Z}, \cite{2009ApJ...698.1122O}, \cite{2010A&A...513A..79B}.  Due to the fact that circumstellar disks exhibit a wide range of temperatures, they radiate from micron wavelengths to millimeter  wavelengths, which is why  they can be observed with infrared and radio telescopes. With the construction of different kinds of these  telescopes, e.g. Spitzer, Herschel, SMA, EVLA or  ALMA,  astronomers can observe with more details the material inside accretion disks around young stars. The parallel development of theory and observations have allowed astrophysicists  to study the different stages of  planet formation, making this topic  one of the most active fields in astronomy today.

In the first stage of planet formation, the growth from sub-micron sized particles to larger objects is a complex process that contains many physical challenges. In the case of smooth disk with a radial gas pressure profile that is monotonically decreasing, the dust particles drift inwards owing to the fact that the gas moves with sub-keplerian velocity due to the gas pressure gradient. Before a large object can be formed, the radial drift  causes dust pebbles to move towards the star. Moreover, the high relative velocities due to turbulence and radial drift cause  the solid particles to reach velocities  that lead to fragmentation collisions which do not allow dust particles to form larger bodies \citep{1977MNRAS.180...57W, 2008A&A...480..859B, 2010A&A...516L..14B}. The combination of these two problems is called ``meter-size barrier'' because on timescales as short as 100 years, a one meter sized object at 1 AU moves towards the star due to the radial drift, preventing that any larger object could be formed.

The observations in the inner regions of the disk, where planets like Earth should be formed,  are very difficult because these regions are so small on the sky that few telescopes can spatially resolve them. Also, these regions are optically thick. However, what amounts to the meter-size barrier in the inner few AU is a ``millimeter-size barrier'' in the outer regions of the disk. These outer regions ($\gtrsim 50$ AU)  are much easier to spatially resolve and are optically thin. Moreover, one can use millimeter observations, which probe precisely the grain size range of the millimeter-size. Therefore, the study of dust growth in the outer disk regions may teach us something about the formation of planets in the inner disk regions.
 
Observations of protoplanetary disks at sub-millimeter and mm wavelengths show that the disks remain dust-rich during several million years with large particles in the outer regions \citep{2007prpl.conf..767N, 2010A&A...512A..15R}. However, it is still unclear how to prevent the inward drift and how to explain theoretically that mm-sized particles  are observed in the outer regions of the disk.  Different mechanisms of planetesimal formation have been proposed to resist the rapid inwards drift like: gravitational instabilities \citep{2002ApJ...580..494Y}, the presence of zonal flows\citep{2009ApJ...697.1269J, 2011A&A...529A..62J, 2011ApJ...736...85U} or dead zones of viscously accumulated gas which form vortices \citep{2006A&A...446L..13V}.  With the model presented here, we want to imitate mechanisms that allow to have long-lived pressure inhomogeneities in protoplanetary disks, by artificially adding pressure bumps onto a smooth density profile.

%%%%%%%%%%%%
%TABLE 1
%%%%%%%%%%%%
\begin{table}
\caption{Parameters of the model}    
\label{table1}     
\centering                         
\begin{tabular}{c c }       
\hline\hline                 
Parameter & Values \\    
\hline                       
   $A$ & $\{0.1; 0.3; 0.5; 0.7\}$  \\          
   $f$ & $\{0.3; 0.7; 1.0; 3.0\}$  \\                                             
   $\alpha$ & $10^{-3}$  \\   
   $R_{\star} [R_\odot]$ & $2.5$  \\   
   $T_{\star} [K]$ & $4300$  \\   
   $M_{disk} [M_\odot]$ & $0.05$  \\
   $\rho_s[g/cm^3]$& $1.2$  \\          
   $v_f[m/s]$& $10$  \\     
\hline                     
\end{tabular}  
\end{table}

To confront the millimeter-size barrier, it is necessary to stop the radial drift considering a radial gas pressure profile that is not monotonically decreasing with radius. Instead, we take a pressure profile with local maxima adding a sinusoidal perturbation of the density profile. These  perturbations influence directly the pressure, following a simple equation of state for the pressure in the disk. Depending on the size of the particle, the dust grains will be nearly perfectly trapped in the pressure peaks, because a positive pressure gradient can cause those dust particles move outwards. On the other hand, turbulence can mix part of the dust particles out of the bumps, so that overall there may still be some net radial inward drift. More importantly, dust fragmentation may convert part of the large particles into micron size dust particles, which are less easily trapped and thus drift more readily inward. 

In the work of \cite{2010A&A...516L..14B}, they compared the observed fluxes and mm spectral indices from Taurus \citep{2010A&A...512A..15R} and Ophiucus \citep{2010A&A...521A..66R} star-forming regions with predicted fluxes and spectral indices at mm wavelengths. They neglected the radial drift, forcing the dust particles to stay  in the outer disk regions. They aimed to keep the spectral index at low values, which implies that the dust particles could acquire millimeter sizes \citep {1991ApJ...381..250B}. However, they found over-predictions of the fluxes. As an extension of their work, the purpose of this paper is to model the combination of three processes:  the radial drift,  the radial turbulent mixing and  the dust coagulation/fragmentation cycle in a bumpy surface density profile. Our principal aim is to find out how the presence of pressure bumps  can help explain the retainment of dust pebbles in the outer regions of protoplanetary disks, while still allowing for moderate drift and thus obtaning a better match with the observed fluxes and mm spectral indeces. In addition, we show simulated images using different antenna configurations of the complete stage of ALMA, to study if it is possible to detect these kind of inhomogeneities  with future ALMA observations.

This paper is ordered as follows: Sect. \ref{sec2} will describe the coagulation/fragmentation model and the sinusoidal perturbation that we take for the initial condition of the gas surface density. Section \ref{sec3}  will describe the results of these simulations, the comparison between existing mm observations of young forming disks and the results from our model.  We discuss  if the type of structures generated by our model can be detectable with  future  ALMA observations. In Sect. \ref{sec4}, we explore  the relation of our model with  predictions of current simulations of the magnetorotational instability (MRI) \citep{1991ApJ...376..214B}. Finally, Sect. \ref{sec5} will summarize our results and the conclusions of this work.

%%%%%%%%%%%%
%FIGURE 1
%%%%%%%%%%%%
 \begin{figure*}
   \centering
   \includegraphics[width=18cm]{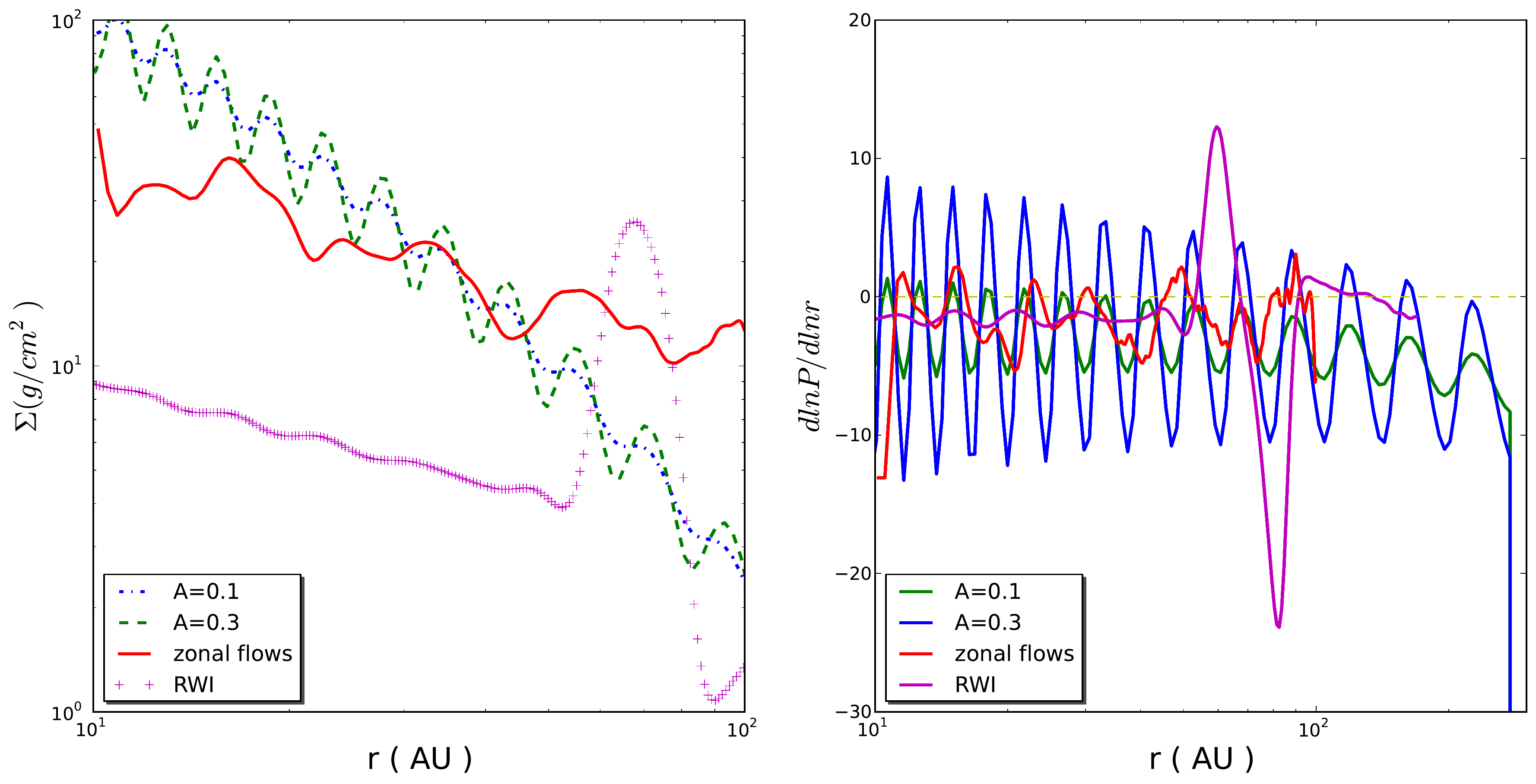}
   \caption{Comparison between: The gas surface density (left plot) taken in this work (Eq. \ref{eq1}) for two different values of the amplitude and constant width (dashed and dot-dashed lines). The Rossby wave instability \citep{2011arXiv1109.6177R}, and the presence of zonal flows due to MHD instabilities  \citep{2011ApJ...736...85U}. Right plot shows the pressure gradient for each of the gas surface density profiles.}
   \label{comparacion}
\end{figure*}

%%%%%%%%%%%%%
\section{Dust Evolution Model} \label{sec2}
%%%%%%%%%%%%%

We make use of the model presented in \cite{2010A&A...513A..79B} to calculate the evolution of dust surface density in a  gaseous disk, radial drift, and turbulent mixing. The dust size distribution evolves due to grain growth, cratering and fragmentation. Relative velocities due to Brownian motion, turbulence, radial and azimuthal drift as well as vertical settling are taken into account.

In this work, we do not consider the viscous evolution of the gas disk because the aim is to investigate how dust evolution is influenced by stationary perturbations of an otherwise smooth gas surface density. The effects of time dependent perturbation and the evolution of the gas disk will be the subject of future work. For a comprehensive description of the numerical code we refer to  \cite{2010A&A...513A..79B}.

In order to simulate radial pressure maxima that allow the trapping of particles, we consider a perturbation of the gas surface density  that it is taken for simplicity as a sinusoidal perturbation such that:   

\begin{equation}
	\Sigma'(r)=\Sigma(r)\left(1+A\cos{\left[ 2\pi \frac{r}{L(r)}\right]}\right),
	\label{eq1}
\end{equation}

\noindent where the unperturbed gas surface density  $\Sigma(r)$ is given by the self similar solution of \cite{1974MNRAS.168..603L}: 

\begin{equation}
	\Sigma(r)=\Sigma_0 \left(\frac{r}{r_c}\right)^{-\gamma}  \exp \left[-\left(\frac{r}{r_c}\right)^{2-\gamma}\right],
	\label{eq2}
\end{equation}

\noindent where $r_c$ is the characteristic radius, taken to be $60$AU, and $\gamma$ is the viscosity power index equal to $1$, which are  the median values found from high angular resolution imaging in the sub-mm of disks in the  Ophiucus star forming regions \citep{2010ApJ...723.1241A}. The wavelength $L(r)$ of the sinusoidal perturbation depends on vertical disk scale-height $H(r)$  by a factor $f$ as

\begin{equation}
	L(r)=f H(r), 
  	\label{eq3}
\end{equation}
   
\noindent with $H(r)= c_s  \Omega^{-1}$, where the isothermal sound speed $c_s$ is defined as

\begin{equation}
	 c_s^2=\frac{k_B T}{\mu m_p}, 
  	\label{eq3-1}
\end{equation}

\noindent  and the Keplerian angular velocity $\Omega$ is

\begin{equation}
	 \Omega=\sqrt{\frac{GM_\star}{r^3}},
  	\label{eq3-2}
\end{equation}

\noindent with $k_B$ being the Boltzmann constant, $m_p$ the mass of the proton and $\mu$ is the mean molecular mass, which in proton mass units is taken as $\mu=2.3$. For an ideal gas, the pressure is defined as
   
\begin{equation}
	P ( r , z )=c_s^2 \rho( r, z ),
  	\label{eq4}
\end{equation}

	\noindent where $\rho(r,z)$ is the gas density, such that $\Sigma'(r)=\int_{-\infty}^\infty \rho(r,z) dz$. With the surface density described by Eq. \ref{eq1}, we can have pressure bumps such that the wavelength is increasing with radius. These bumps are static, which may not be entirely realistic. However, these can be a good approximation of long-lived, azimuthally extended pressure bumps, that can be e.g. the result of  MHD effects \citep[see][]{2009ApJ...697.1269J, 2010A&A...515A..70D}. The influence of time-dependent pressure perturbations \citep[e.g.][]{2004ApJ...608..489L, 2007Icar..188..522O} on the dust growth process will be the topic of future work. Left plot of Fig. \ref{comparacion} (dashed lines) shows the behavior of the perturbed surface density for two values of the amplitude and fixed value of the width. The right plot of Fig. \ref{comparacion} shows the corresponding pressure gradient.

The very fine dust particles move as the gas  because they are well coupled to the gas since the stopping time is very short. In the presence of a drag force, the stopping time is defined as the time that a particle, with a certain momentum, needs to be aligned to the gas velocity. However, when the particles are large enough and they are not forced to move as the gas, they experience a  head wind, because of the sub-Keplerian velocity of the gas and therefore they lose angular momentum and move inwards. In this case, the resulting drift velocity of the particles is given by  \cite{1977MNRAS.180...57W}:

\begin{equation}
	u_{\mathrm{drift}}=\frac{1}{\textrm{St}^{-1}+\textrm{St}} \frac{\partial_r P}{\rho \Omega}. 
  	\label{eq5} 
\end{equation}

Comparing Eq. \ref{eq5} with the expression for the drift velocity given by \cite{2010A&A...513A..79B} (Eq. 19), we can notice that the drag term from  the radial movement of the gas is not taken here since we are assuming a stationary gas surface density. The Stokes number denoted by St,  describes the coupling of the particle to the gas. The Stokes number is defined as the ratio between  the eddy turn-over time ($1/\Omega$) and  the stopping time. For larger bodies, the Stokes number is much greater than one, which implies that the particles are not affected by the gas drag, consequently they move on Keplerian orbits. When the particles are very  small, St $\ll$ 1, they are strongly coupled to the gas. Since the gas is orbiting at sub-Keplerian velocity because its pressure support, there is a relative velocity between the dust particles and the gas. The Stokes number equal to unity is a critical value where the particles are still influenced by the gas drag but they are not completely coupled to the gas, instead they are marginally coupled, and move at speeds between Keplerian and the sub-Keplerian gas velocity.

The retainment of dust particles due to the presence of pressure bumps depend on the size of the particles. Since very small particles, with St $\ll$ 1, are tightly coupled to the gas, they do not drift inwards. However, radial drift becomes important when the size of the particles increases and it is strongest when St=1 (see Eq. \ref{eq5}). In the Epstein regime, where the ratio between the mean free path of the gas molecules $\lambda_{\mathrm{mfp}}$ and the particle size, denoted by $a$,  satisfies that $\lambda_{\mathrm{mfp}}/a\geq 4/9$, the Stokes number is given by:

\begin{equation}
	\textrm{St}=\frac{a\rho_s}{\Sigma_g}\frac{\pi}{2},
  	\label{eq6}
\end{equation}

\noindent where $\rho_s$ is the solid density of the dust grains, that is taken to be constant (see Table \ref{table1}). In this case particles are small enough to be in this regime.  Parameterizing the radial variation of the sound speed via 

\begin{equation}
	c_s\propto r^{-q/2},
  \label{eq7}
\end{equation}

\noindent where for a typical disk, the temperature is assumed to be a power law such that $T\propto r^{-q}$, which is an approximation of the temperature profile taken for this model. Therefore, the wavelength of the perturbation $L(r)$ scales as:

\begin{equation}
	L(r)=f H(r)\propto f r^{(-q+3)/2}.
  \label{eq8}
\end{equation}
 
The pressure bumps have the same amplitude $A$ and wavelength $L(r)$ than the density, because the over pressure is induced adding inhomogeneities to the gas surface density and parameterizing the temperature on the midplane by a power law \citep{1994ApJ...421..640N}. The model taken here can artificially imitate e.g. the case of  zonal flows in protoplanetary disks, where over densities create pressure bumps. Zonal flows are formed due to MRI, which depend on the degree of ionization of the disk, i.e. on the temperature of the disk and other factors as the exposure to cosmic and stellar rays. MRI appears to be the most probable source of turbulence, and if the turbulence is not uniform, there can be excitation of long-lived pressure fluctuations in the radial direction. For instance, \cite{2009ApJ...697.1269J} performed shearing box simulations of MRI turbulent disk that show large scale radial variations in Maxwell stresses of $10\%$. \cite{2010A&A...515A..70D} presented 3D global non-ideal MHD simulations including a dead zone that induces pressure maxima of $20-25\%$. \cite{2011ApJ...736...85U} showed 3D global MHD simulations, leading to pressure bumps of  around $25\%$. On the other hand, when the viscosity drops, the gas surface density changes  causing a local inversion of the pressure gradient and an accumulation of dust particles. This matter accumulation causes Rossby wave instabilities \citep{1999ApJ...513..805L} that lead non-axisymmetric distributions on the disk, which we cannot exactly model at this moment since our dust evolution models are axisymmetric. 

To constrain the values of the amplitude and wavelength of the perturbation, we take into account three different factors: First, it is important to analyze the necessary conditions to have local outwards movement of the dust particles. Second, we compare our assumptions with current studies on zonal flows \citep{2009ApJ...697.1269J, 2011ApJ...736...85U}. And third, we only work with small-enough amplitude disturbances such that the disk has angular momentum per unit mass increasing outwards, which means it is Rayleigh stable . 

The Rayleigh criterion establishes that for a rotating fluid system, in the absence of viscosity, the arrangement of angular momentum 
per unit mass is stable if and only if it increases monotonically outward \citep{1961hhs..book.....C}, which implies:

\begin{equation}
\frac{\partial }{\partial r}(r v_{\phi})>0.
  \label{eq16}
\end{equation}

%%%%%%%%%%%%
%FIGURE 2
%%%%%%%%%%%%
 \begin{figure}
   \centering
   \includegraphics[width=8.8cm]{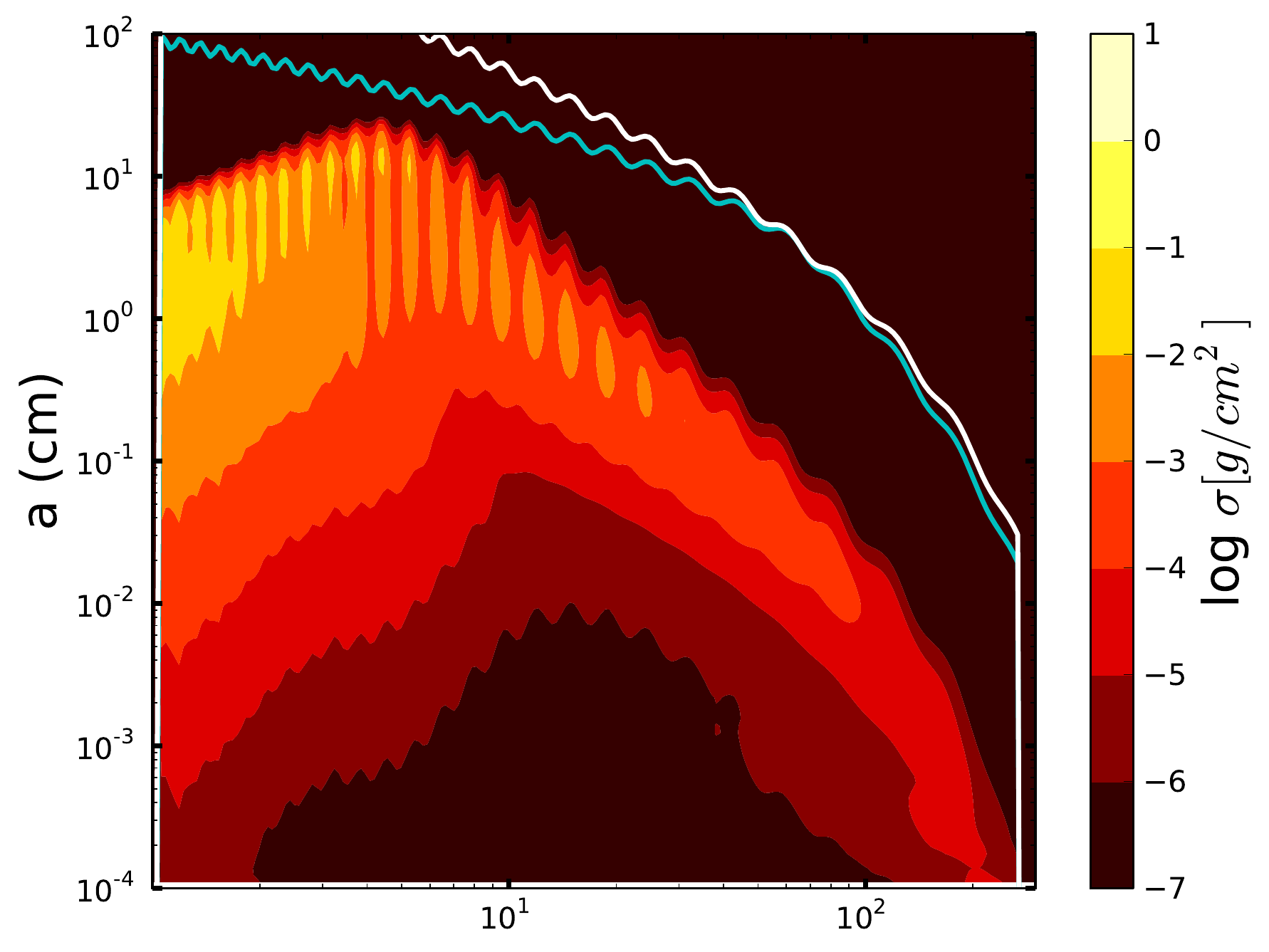}\\
   \includegraphics[width=8.8cm]{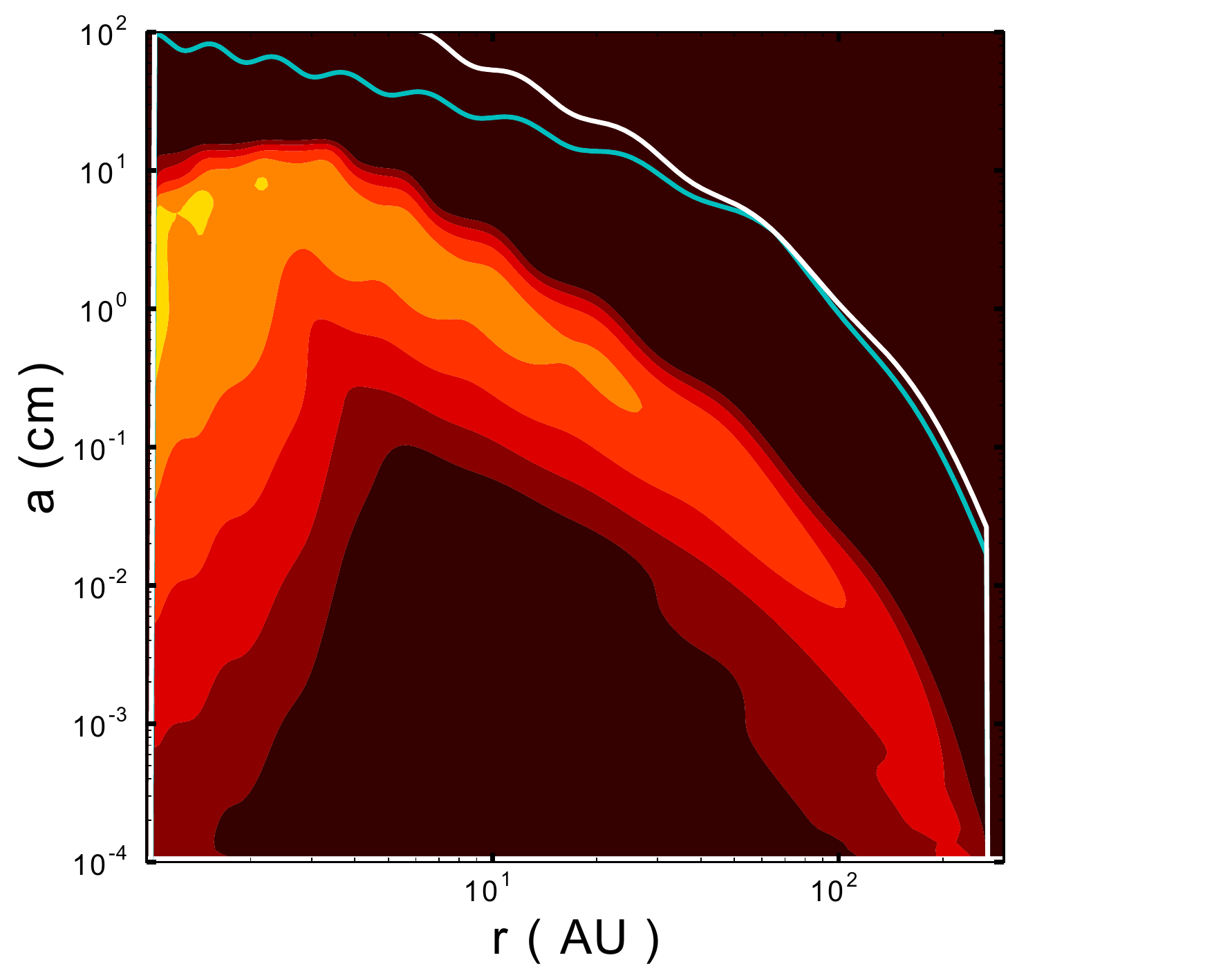}
   \caption{Vertically integrated dust density distribution  at 1 Myr for $A=0.1$ and $f = 1$ (top) and $A=0.1$ and $f=3$ (bottom). The  white line shows the particle size corresponding to a Stokes number of unity, which shows the same shape that the gas surface density $\Sigma'$ of Eq. \ref{eq1} (see Eq. \ref{eq6}). The blue line represents the maximum size of the particles before they reach fragmentation velocities (fragmentation barrier according to Eq.\ref{eq9}).}
   \label{Fig1}
\end{figure}

Since turbulence is necessary to have angular momentum transport,  instabilities may occur if a magnetic field is present, and in that cases the angular velocity decreases with radius (MRI). For a typical $\alpha-$turbulent disk, the MRI time scale is much greater than the time that the disk needs to recover the Rayleigh stability, this implies that the disk should remain quasi stable at all time \citep{2010MNRAS.402.2436Y}. Any perturbation in the gas surface density, that is Rayleigh unstable will almost instantly be smeared out sufficiently to make it Rayleigh stable again, thereby lowering its amplitude. This happens on a time scale much shorter than what MRI could ever counteract. The angular velocity of Eq. \ref{eq16} is given by:  

\begin{equation}
v_{\phi}^2=v_k^2+\frac{r}{\rho}\frac{\partial P}{\partial r}=v_k^2(1-2\eta) 
  \label{eq17}
\end{equation}

\noindent  with 

\begin{equation}
 \eta=-\frac{1}{2r\Omega^2 \rho}\frac{\partial P}{\partial r}.
 \label{eq17.0}
\end{equation}

%%%%%%%%%%%%
%FIGURE 3
%%%%%%%%%%%%
 \begin{figure*}
   \centering
   \includegraphics[width=18cm]{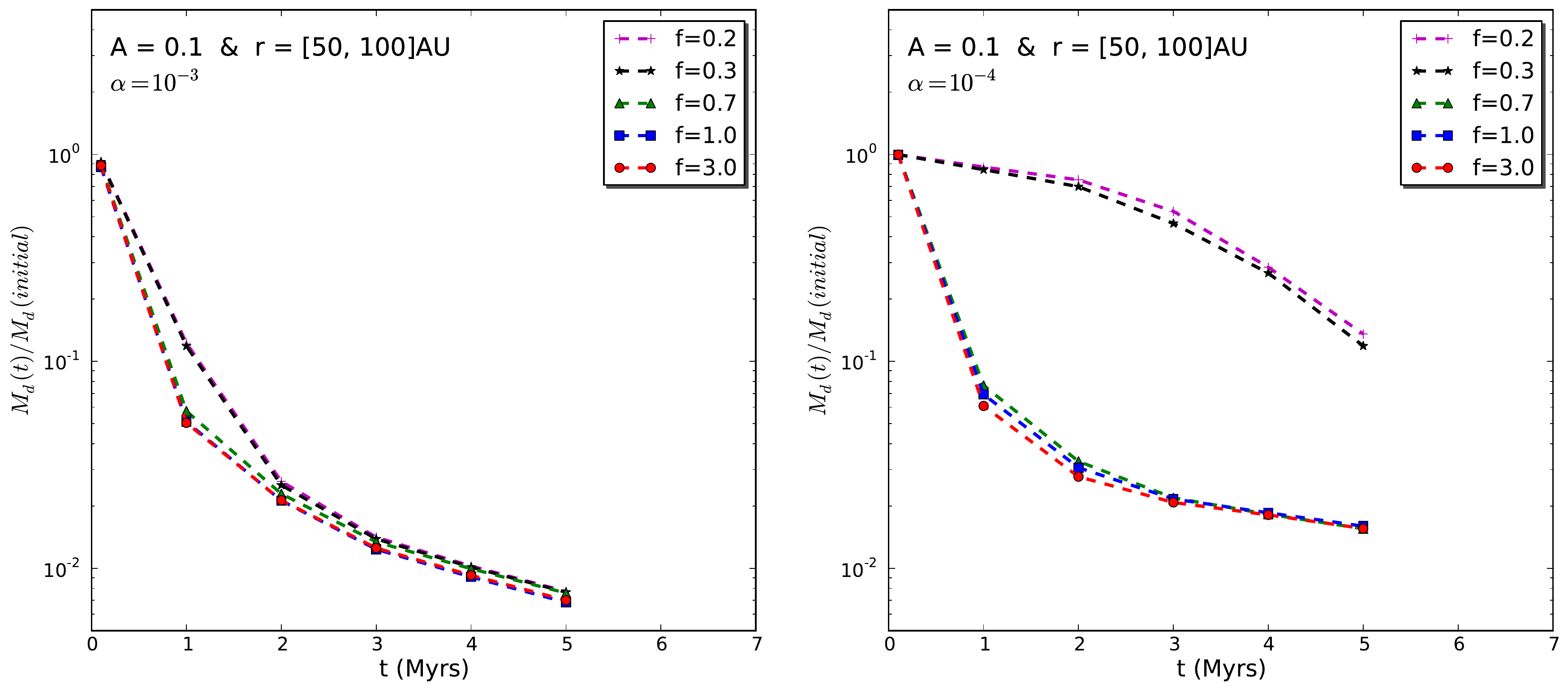}
   \caption{Ratio between final and the initial dust mass between $50$ AU  and $100$AU, at different times of evolution. Taking a constant value of the amplitude $A=0.1$ and different values of the width of the perturbed density (Eq. \ref{eq1}). For $\alpha=10^{-3}$ left plot and $\alpha=10^{-4}$ right plot}
   \label{width}
\end{figure*}

The Rayleigh stability of the disk depends on the amplitude and the width of the bumps. In this case, we want to study the influence of the amplitude of the perturbation, so for this analysis we constrain the value of the wavelength of the perturbation, $f$ equals to unity, such that it stays consistent with the values expected from predictions of zonal flows models by \cite{2011ApJ...736...85U}  (see Fig. \ref{comparacion}). 

Taking the perturbed gas density of Eq. \ref{eq1} and $f=1.0$, it is possible to find the upper limit of the amplitude to satisfy the Eq. \ref{eq16}, i.e. the condition to remain Rayleigh stable at all time. This calculation lets the maximum value of the amplitude $A$  to be at most  $\sim 35\%$ of the unperturbed density. 

Various investigations have been focused on the possibility of Rayleigh instabilities when disks have sharp profiles for the radial density (see e.g. \cite{1984ApJ...285..818P}, \cite{2001ApJ...551..874L} and \cite{2010MNRAS.402.2436Y}). These kind of profiles can exist when the temperature in the midplane is not sufficient to ionize the gas \citep{1996ApJ...457..355G} and as a result the turbulence parameter $\alpha$ is reduced. These regions are known as  ``dead zones'' and these are candidates to be places where planet embryos can be formed. In the boundaries of these regions, it was shown by \cite{2006A&A...446L..13V}  that it is possible to have a huge vortex with a local bump in the gas surface density. \cite{1999ApJ...513..805L} demonstrated that  these perturbations create an accumulation of gas that leads to the disk  to be unstable to Rossby wave instability (RWI). As a comparison of the amplitudes generated by  RWI vortices and the amplitudes of our perturbations, the left plot of Fig. \ref{comparacion} also shows the azimuthally average gas surface density of a large-scale anticyclonic vortex by \cite{2011arXiv1109.6177R}. In those cases, the equilibrium of the disk is affected such that the disk may become Rayleigh unstable. Since we are focused on perturbations that allow to the disks stay Rayleigh stable at all evolution time, we do not consider for our perturbed density such kind of amplitudes.

On the other hand, since the drift velocity is given by Eq. \ref{eq5},  to prevent the inwards drift, the value of $\eta$ from Eq. \ref{eq17.0} must be negative, this implies that the pressure gradient has to be positive for some regions of the disk.  Doing this calculation for the condition that $\eta<0$, we get that  the values of the amplitude $A$ have to be at least equal to about $10\%$. In right plot of Fig. \ref{comparacion} we see that with an amplitude of 10\% the pressure gradient barely reaches positive values in the inner regions of the disk ($\lesssim 50$ AU). Summarizing the upper and lower values of the amplitude  should be $0.10\lesssim A \lesssim 0.35$, when the width of the perturbation is taken to be $f=1$.

Taking into account the growth-fragmentation cycle and the existence of pressure bumps, the radial drift efficiency can be reduced if the bumps have  favorable values for the amplitude and the length scale.  When the particles grow  by coagulation, they reach a certain size with velocities high enough to cause fragmentation (fragmentation barrier). The two main sources of relative velocities are radial drift and turbulence. In the bumps the radial drift   can be zero if the pressure gradient is high enough as well as azimuthal relative velocities; but there are still relative velocities due to the turbulence. Therefore, it is necessary to have a condition, such that the particles do not reach the threshold where they fragment. The maximum turbulent relative velocity between particles, with St $\sim1$, is given by \cite{2007A&A...466..413O},  

\begin{equation}
	\Delta u_{max}^2\simeq \frac{3}{2}\alpha  \mathrm{St} c_s^2,
  \label{eq12}
\end{equation}
 
\noindent  for St $\lesssim0.1$  Eq.\ref{eq12} is off by a factor of 2. Therefore, to break through the mm size barrier, we must have that $\Delta u_\mathrm{max}$ has to be smaller that the fragmentation velocities of the particles $v_f$. Recent collision experiments using silicates \citep{2010A&A...513A..56G} and numerical simulations \citep{2010A&A...513A..57Z} show that there is an intermediate regime between fragmentation and sticking, where particles should bounce. In this work, we do not take into account this regime since there are still many open questions in this field. For example, \cite{2011ApJ...737...36W} suggest that there is no bouncing regime for ice particles, which may be present in the outer regions of the disks  \citep[see][]{2010A&A...517A..87S}. Laboratory experiments and theoretical work suggest that typical values for fragmentation velocities are of the order of few m s$^{-1}$ for silicate dust \citep[see e. g.][]{2008ARA&A..46...21B}. Outside the snow-line, the presence of ices affects the material properties, making the fragmentation velocities increase \citep{2007A&A...470..733S, 2009ApJ...702.1490W}. Since in this work we assume a radial range from 1AU to 300AU, the fragmentation threshold velocity is taken as $v_f=10$m s$^{-1}$. All the parameters used for this model are summarized in Table ~\ref{table1}. 

For particles with $ \mathrm{St} \lesssim 1$, taking the size at which the turbulent relative velocities $\Delta u_\mathrm{max}$ are as high as the fragmentation velocity $v_f$, we can find the maximum value of the grain size, which is 

\begin{equation}
	a_{\mathrm{max}}\simeq\frac{4\Sigma_g}{3\pi \alpha\rho_s} \frac{v_f^2}{c_s^2},
  \label{eq9}
\end{equation}

\noindent this $a_{\mathrm{max}}$ is valid only for $ \mathrm{St} \lesssim 1$ because for larger bodies the turbulent relative velocities are lower than the given in Eq. \ref{eq12} (for a detailed discussion about turbulent relative velocities see  \cite{2007A&A...466..413O}).

The dust grain distribution $n(r,z,a)$ is the number of particles per cubic centimeter  per gram interval in particle mass, which depends on the grain mass, distance to the star $r$ and height above the mid-plane $z$, such that

\begin{equation}
	\rho (r,z)=\int_0^\infty n(r,z,a) \cdot m dm,
  \label{eq10}
\end{equation}

%%%%%%%%%%%%
%FIGURE 4
%%%%%%%%%%%%
 \begin{figure}
  \centering
  \includegraphics[width=8.8cm]{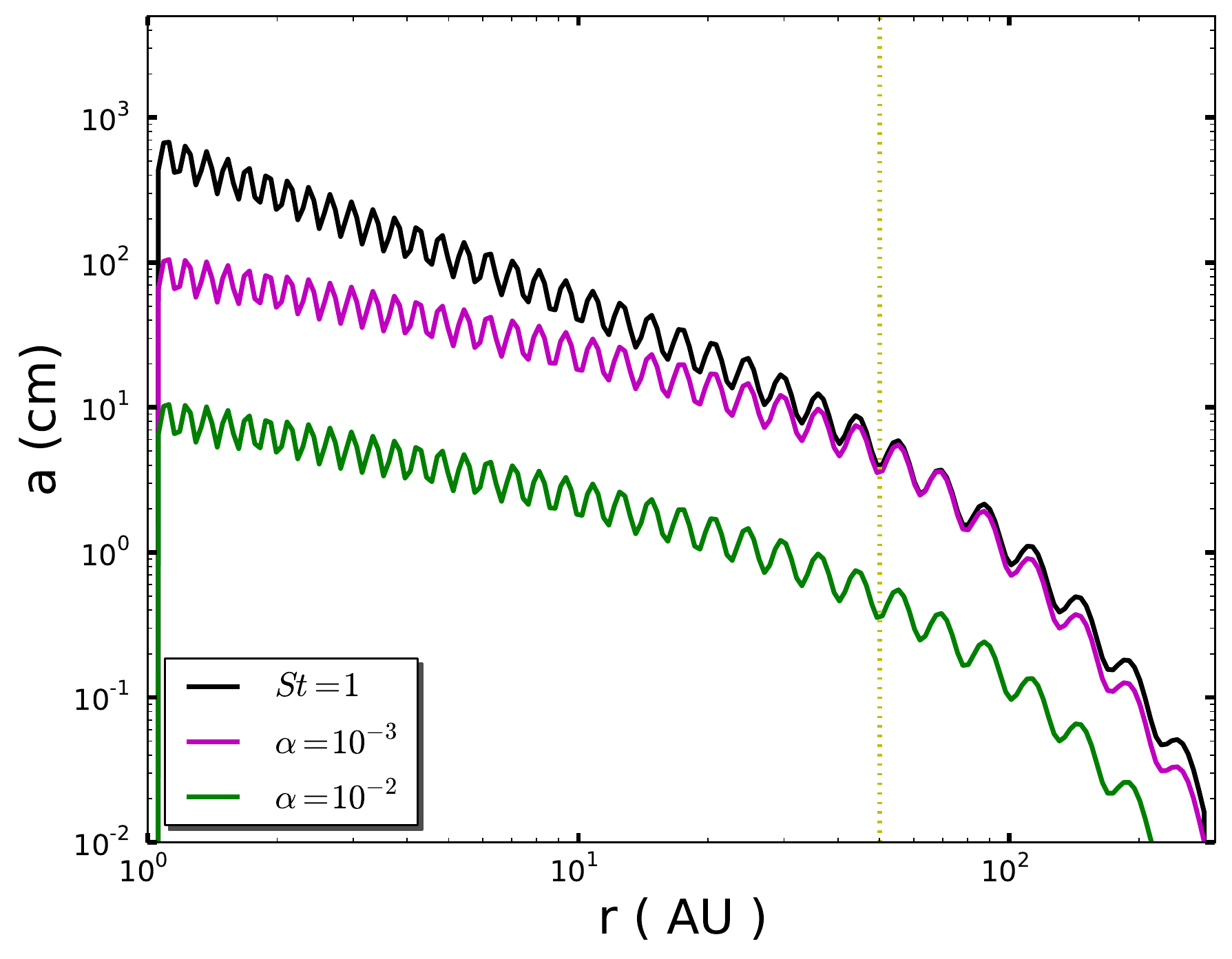}
   \caption{Particle size corresponding to a Stokes number of unity for $A=0.3$ and $f=1.0$ and location of the fragmentation barrier for two different values of the turbulent parameter $\alpha$. The dashed line corresponds to $r=50$AU to distinguish the maximum size particle in the outer regions of the disk for each case.}
   \label{comparison_alpha}
 \end{figure}

\noindent is the total dust volume density. The quantity $ n(r,z,a)$ can change due to grain growth and distribution of masses via fragmentation. The vertically integrated dust surface density distribution per logarithmic bin defined as

\begin{equation}
	\sigma (r,a)=\int_{-\infty}^{\infty} n(r,z,a)\cdot m\cdot a dz
  \label{eq11}
\end{equation}

\noindent and the total dust surface density is then

\begin{equation}
	\Sigma_d(r)=\int_0^\infty \sigma (r,a)d\ln a.
  \label{eq11}
\end{equation}

%%%%%%%%%%%%%
\section{Results} \label{sec3}
%%%%%%%%%%%%%

%%%%%%%%%%%%%
\subsection{Density distribution of dust particles} \label{subsec3.1}
%%%%%%%%%%%%%

%%%%%%%%%%%%
%FIGURE 5
%%%%%%%%%%%%
 \begin{figure*}
  \centering
  \includegraphics[width=18.0cm]{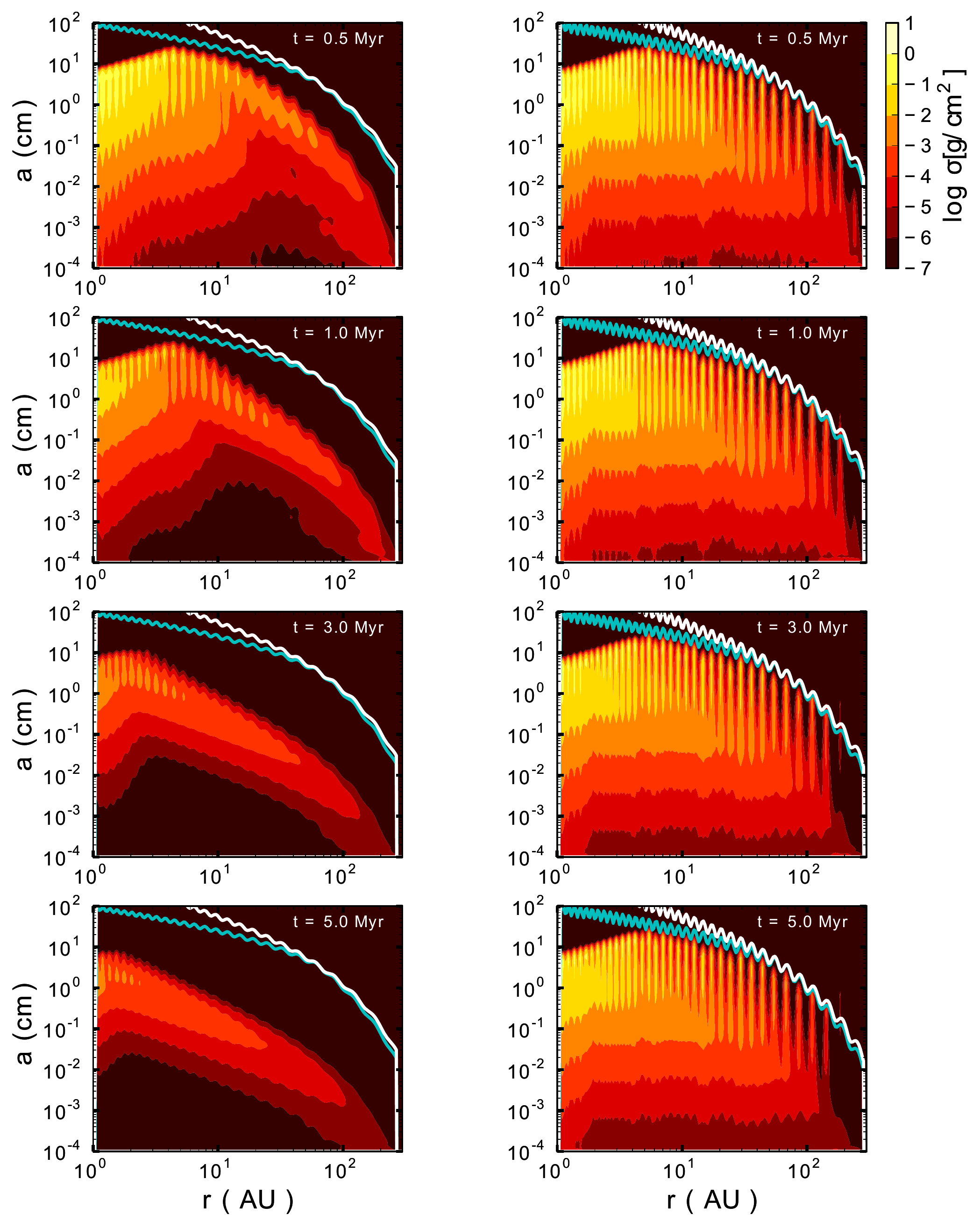}
   \caption{Vertically integrated dust density distribution with fixed value of length scale as $f = 1$, for $A=0.1$ (left column) and $A=0.3$ (right column) at different times $0.5$ Myr, $1.0$ Myr, $3.0$ Myr and $5.0$ Myr from top to bottom respectively. The solid white line shows the particle size corresponding to a Stokes number of unity. The blue line represents the fragmentation barrier according to Eq. \ref{eq9}.}
   \label{Fig2}
 \end{figure*}
 
The simulations have been done with a disk of mass $0.05 M_\odot$, with a surface density described by Eq. \ref{eq1} from $1.0$ AU to $300$ AU, around a star with one solar mass. The turbulence parameter $\alpha$ is taken to be $10^{-3}$, unless other value is specified. Figure \ref{Fig1} shows the  vertically integrated dust density distribution,  taking into account: coagulation, radial mixing, radial drift and fragmentation, after $1$ Myr of the evolution of the protoplanetary disk. The solid white line shows the particle size corresponding to a Stokes number of unity. From Eq. \ref{eq6} we can see that when St=1, the particle size $a$ is proportional to the gas surface density $\Sigma '$, then the solid line reflects the shape of the surface density. The blue line of Fig. \ref{Fig1} represents the fragmentation barrier, which illustrates the maximum size of the particles  before they reach velocities higher than the  fragmentation velocity (see  Eq. \ref{eq9}). Hence, particles above the fragmentation barrier should fragment down to smaller particles, which again contribute to the growth process.   

Both plots of Fig. \ref{Fig1} have the same amplitude of the sinusoidal perturbation $A=0.1$. The factor $f$ which describes the width of the perturbation, is taken to be $f=1$ for the top plot,  and $f=3$ for the bottom plot of Fig. \ref{Fig1}. This result shows that: First, the amplitude $A=0.1$ of the perturbation is not high enough to have a positive pressure gradient in those regions (see right plot of Fig. \ref{comparacion}) such that  particles can be retained in the outer regions of the disk after some Myr. Instead the dust particles are still affected by radial drift and turbulence such that the particles do not grow over mm size in the outer regions. 

Second, taking a greater value of the factor $f$, at the same amplitude, implies that the retention of particles is even weaker. This is because with a wider perturbation is it harder to have positive pressure gradient. It is expected that for a smaller value of $f$, the pressure gradient is higher since the surface profile should be steeper and therefore a dust trapping much more  efficient. However, the diffusion timescale $\tau_{\nu}$ depends quadratically on the length scale $\ell$. Therefore, when the  wiggles of the perturbation are taken with a smaller wavelength, the diffusion times become much shorter, implying that  the turbulence mixes the dust particles out of the bumps faster, even when the pressure gradient is higher for narrow wiggles. More precisely, $\tau_{\nu}\propto \ell^2 \nu^{-1} $ where the viscosity is defined as $\nu = \alpha c_s h$. For this reason, we notice in Fig. \ref{width}-left plot that the trapping is  more effective taking values of the width less than one, but only  by a very small margin. As a result, the ratio between the final and the initial mass of the dust for $r \in [50, 100] $AU remains almost constant when the width is taken larger than $0.3$. We conclude that for an amplitude of $A$=0.1, the trapping after several million years, does not become more effective when the wavelength of the perturbation is taken shorter.

Only when the diffusion timescales become larger or equal to the drift timescales for a given pressure profile, a turbulence parameter and a Stokes number,  the dust particles can be retained inside the bumps and therefore they can grow. From Eq. \ref{eq5} we can deduce that the drift time scales as $\tau_{\mathrm{drift}}\propto \ell (\partial_{\ell} P)^{-1}$, where inside the bumps $\ell$ is the width of the perturbations (which depends directly on $f$).  As a consequence, after an equilibrium state is reached, drift and diffusion timescales are both proportional to the square of the width.  Accordingly, for a given value of $f$, the effect of turbulent mixing and radial drift offset.

We can notice in Fig. \ref{width} that for $f$=\{0.3, 0.2\} there is a small effect in the retention of particles for $\alpha=10^{-3}$ (left plot) and an important effect for $\alpha=10^{-4}$ (right plot). This implies that for these values of $f$ and $A$, the pressure gradient becomes positive enough in the outer regions ($r \in [50, 100] $AU) to have trapping of particles. Since $A$ and $f$ are the same for both cases, the pressure gradient is exactly the same. However, for $\alpha=10^{-3}$, the small efficiency that becomes visible reducing $f$, vanishes after two Myr, because turbulent mixing and radial drift cancel each other.

Nevertheless, when $\alpha$ is reduced one order of magnitude (right panel of  Fig. \ref{width}), the diffusion timescales are now longer. In this case, we have  that the drift timescales are shorter than the diffusion timescales, hence the ratio between the final and the initial dust mass increase in average for each $f$.  When $f$ is small enough to have positive pressure gradient ($f$=\{0.3, 0.2\}), outward drift wins over turbulent mixing, and as a result there is a visible effectiveness in the trapping of particles. However, there is almost no difference between $f$=0.3 and $f$=0.2. Contrary, this counterbalance effect between radial drift and turbulence when $f$ varies does not happen when the amplitude of the perturbation changes.
 
We fixed the value of the width of the perturbation to unity, because this value is consistent with current model predictions of zonal flows. The comparison between our assumption of the density inhomogeneities and the work from \citet{2011ApJ...736...85U} is discussed in Sect. \ref{sec4}.  In addition, for longer values of the width, we should have higher values of the amplitudes in order to have a positive pressure gradient. In that case, however the disk becomes easily Rayleigh unstable when the amplitude in increased. These are the reasons why we fix the value of the width to unity and not higher.

Simulations of MRI-active disks suggest that the typical values for the turbulence parameter $\alpha$ are in the range of $10^{-3}-10^{-2}$  \citep {2005ApJ...634.1353J, 2010A&A...515A..70D}. In this work, we focus on the results for $\alpha=10^{-3}$, because with a larger value of the turbulence the viscous time scales become shorter compared with the growth  time scales of the dust, making the particles mix out of the bumps and then drift radially inwards before any mm-sizes are reached. In addition, if $\alpha$ is taken one order of magnitude higher, the fragmentation barrier is lower by about one order of magnitude in grain size, implying that particles do not grow over mm-sizes in the outer regions of the disk (see Eq. \ref{eq12}). Fig. \ref{comparison_alpha} shows the location of the fragmentation barrier for the case of $A=0.3$  and $f=1.0$ and two different values of $\alpha$. We can notice that the maximum value of the grain size for the case of $\alpha=10^{-2}$ is of the order of few mm in the outer regions of the disk $r>50$AU, while for $\alpha=10^{-3}$ the grains even reach cm-sizes.
 
Figure \ref{Fig2} compares the surface density distribution for two different values of  the amplitude of the perturbation $A=0.1$ and $A=0.3$, at different times of evolution. Taking $A=0.3$, we can notice that in the pressure bumps there is high density of dust particles, even after $5$ Myr of evolution for a maximum radius around $80$ AU. For $A=0.3$ and $r\gtrsim 100$AU, there is a small amount of particles above the fragmentation barrier for the different times of evolution. It is important to notice that the line of the fragmentation barrier (\ref{eq9}) is calculated taking into account only turbulent relative velocities, since radial and azimuthal turbulence relative velocities were assuming zero  at the peaks of the bumps. Particles with St$>1$ are not anymore well coupled to the gas, they are not affected by gas turbulence, then relative velocities due to turbulence are lower, which implies that they can grow over the fragmentation barrier.  Moreover,  we can see in the right plot of Fig. \ref{comparacion}, that the pressure gradient for  $A=0.3$ and $r\gtrsim 100$AU is always negative, hence for those regions the pressure bumps may not reduce the efficacy of radial drift. Therefore, the total relative velocities for $r\gtrsim 100$AU can be go down, leading some particles (with St$>1$) to grow  over the fragmentation barrier. Only the particles with St$<1$ and over the fragmentation barrier should eventually fragment down to smaller particles. For regions  $r\lesssim 100$AU, dust particles continuously grow to mm-size particles by coagulation because collision velocities due to turbulence are lower that the taken fragmentation velocity $v_f$.

%%%%%%%%%%%%
%FIGURE 6
%%%%%%%%%%%%
 \begin{figure}
  \centering
  \includegraphics[width=8.8cm]{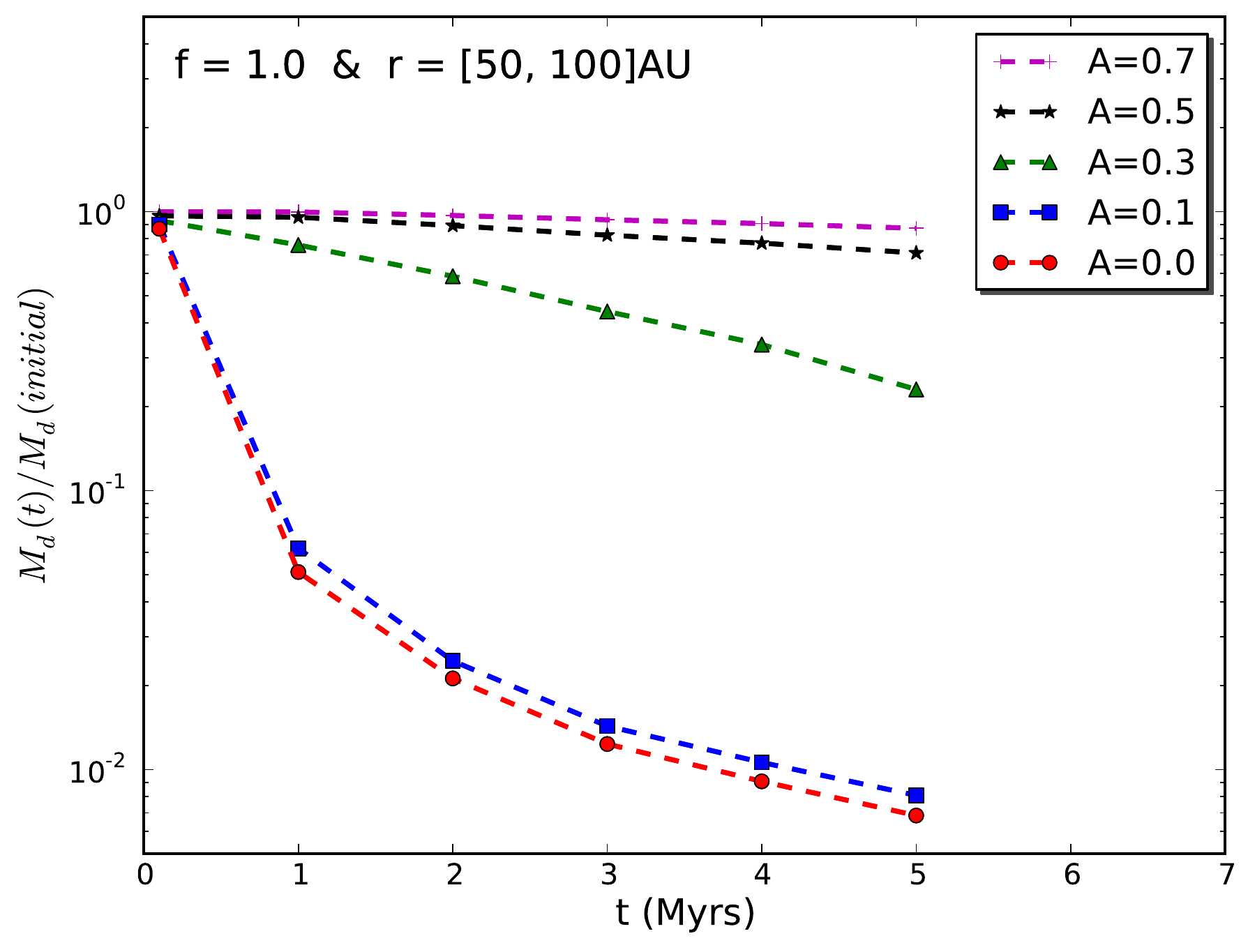}
   \caption{Ratio between final and the initial dust mass between $50$ AU and $100$ AU, at different times of evolution. Taking a constant value of the width $f=1.0$ and different values of the amplitude of the perturbed density (Eq. \ref{eq1}).}
   \label{amplitude}
 \end{figure}

As we mentioned before, the efficiency of the dust trapping will depend on the amplitude of the pressure bumps. It is expected that for higher amplitude there is more trapping of particles, since the pressure gradient is also higher and positive (see right plot of Fig. \ref{comparacion}). Taking the perturbed density of Eq. \ref{eq1}, we can see in Fig. {\ref{amplitude}} that between 50 AU and 100 AU  from the star, the amount of dust grows significantly from $A=0.1$ and $A=0.3$. From $A=0.3$ to $A=0.5$, there is still a considerably growth, but the rate of growth is slower. From $A=0.5$ to $A=0.7$ the rate remains almost constant, reaching a threshold. When the amplitude is increased, the amount of dust particles retained in the bumps rises to a limit when the dust growth is stopped due to the fact that the particles reach the maximum possible value where they start to fragment due to the high relative velocities.

%%%%%%%%%%%%
%FIGURE 7
%%%%%%%%%%%%
  \begin{figure*}
   \centering
  \includegraphics[width=18cm]{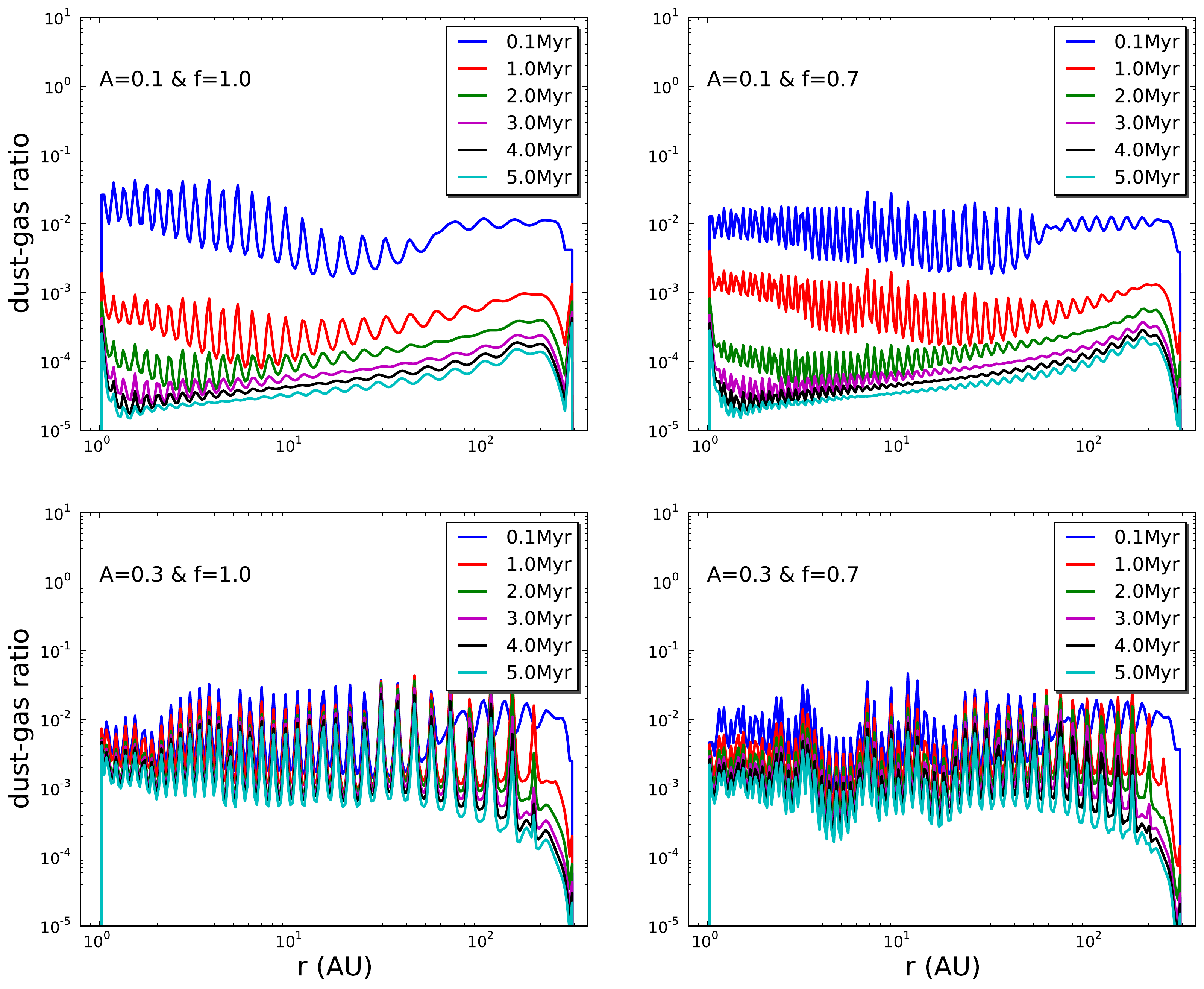}
    \caption{Dust to gas mass ratio at the disk  for different times of disk evolution and the parameters summarizing  in Table \ref{table1}: $A=0.1$ \& $f=1$ (top-left); $A=0.3$ \& $f=1$ (bottom-left); $A=0.1$ \& $f=0.7$ (top-right) and  $A=0.3$ \& $f=0.7$ (bottom-right).}
   \label{ratio}
\end{figure*}

As we explained in Sect. \ref{sec2}, taking a fixed value of $f=1.0$, the disk stays Rayleigh stable for  $0.10\lesssim A \lesssim 0.35$, which means that for those values of the amplitude, these kind  perturbations can be explained via MRI  without any Rayleigh instability present at any evolution time. The amplitude of $A=0.3$ is the one that is more consistent with current MHD simulations of zonal flows \cite{2011ApJ...736...85U}, where pressure reaches radial fluctuations of $25\%$ (see left plot of Fig.  \ref{comparacion}). The amplitude of $A=0.1$ is more consisted with the case of pressure fluctuation of zonal flows of the order of $10\%$ by  \cite{2009ApJ...697.1269J}.

Figure \ref{ratio} shows the radial dependence of the dust-to-gas ratio for different times of the simulation, for two values for the perturbation amplitude and wavelength without the gas inward motion. For $A=0.1$ and $f=1.0$ (top-left plot of Fig. \ref{ratio}), we can see that the dust-to-gas ratio decreased significantly with time in the whole disk.  This implies that the dust particles do not grow considerably with time, which is what we expected due to the fact that with this amplitude the trapping of the particles into the pressure bumps is not effective. Consequently,  due to turbulence the dust particles collide, fragment and become even smaller, so they mix and the retention of those small particles, with St$<1$, becomes more difficult. Thus the radial drift is not reduced and particles  move inwards. Therefore particles with St$\lesssim1$ drift inwards and the particles that survive are those that are very small St$\ll 1$, and are well coupled to the gas. Hence the dust-gas ratio initially decreases quickly and then  becomes almost constant with time, which implies that after several Myr only the very small dust particles remain. Top-right plot of Fig. \ref{ratio} shows that taking the same amplitude but a smaller wavelength, the dust to gas radio has the same behavior. This confirms  that when $\alpha$ turbulence is constant, a decrease of the wavelength implies shorter diffusion time scales. Therefore the trapping is not more effective even if the pressure gradient is higher for narrow bumps.

Conversely, due to the strong over-pressures at $A=0.3$ (bottom plots of Fig. \ref{ratio}), the dust-to-gas ratio remains almost constant with time  for $r< 100$AU, oscillating radially between  $\sim10^{-3}$ to $ \sim10^{-1}$. This oscillating behavior, even after 5 Myr of dust evolution, is possible thanks to the fact that the particles are retained in the bumps and grow enough to make the dust-gas ratio higher inside the bumps. Only around $\sim 100$AU from the star, the dust to gas ratio decreases slowly with time.  This implies that for $r< 100$AU, the drift is counteracted by the positive local pressure gradient, allowing that the time scales for the growth are comparable with the disk evolution times, i. e. with the viscous time scales. Changing the width of the perturbation, $f=1.0$ for the left-bottom plot and $f=0.7$ for the right-bottom plot of Fig. \ref{ratio}, it has just a minor effect over the dust-to gas ratio as was explained before.

%%%%%%%%%%%%%
\subsection{Comparison to current data of young disks in the millimeter range} \label{subsec3.2}
%%%%%%%%%%%%%

In this Section we compare the models predictions of the disk fluxes at millimeter wavelenghts with observational data obtained for young disks in Class II Young Stellar Objects (YSOs).

To do this, we calculate the time-dependent flux for the disk models described in Sect. \ref{sec2}. For the dust emissivity, we adopted the same dust model as in \cite{2010A&A...512A..15R, 2010A&A...521A..66R, 2011A&A...525A..81R} and \cite{2010A&A...516L..14B}, i.e. spherical composite porous grains made of silicates, carbonaceous materials and water ice, with relative abundances from \cite{2003A&A...410..611S}. At each stellocentric radius in the disk, the wavelength-dependent dust emissivity is calculated considering the grain size number density $n (r,z,a)$ derived from the dust evolution models at that radius, as described in Sect. \ref{sec2}. 

%%%%%%%%%%%%
%FIGURE 8
%%%%%%%%%%%%
\begin{figure}
   \centering
  \includegraphics[width=8.8cm]{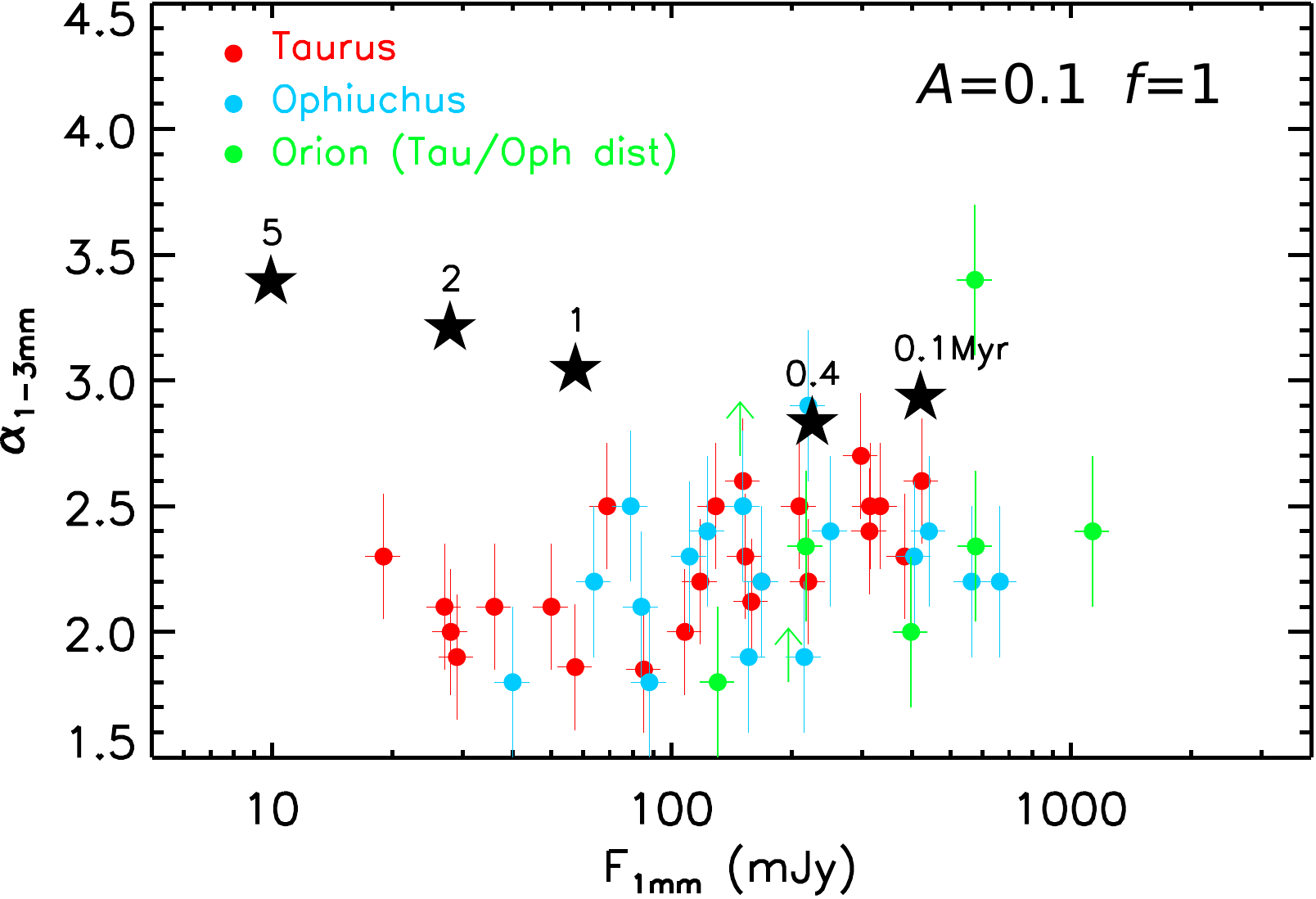}\\
  \includegraphics[width=8.8cm]{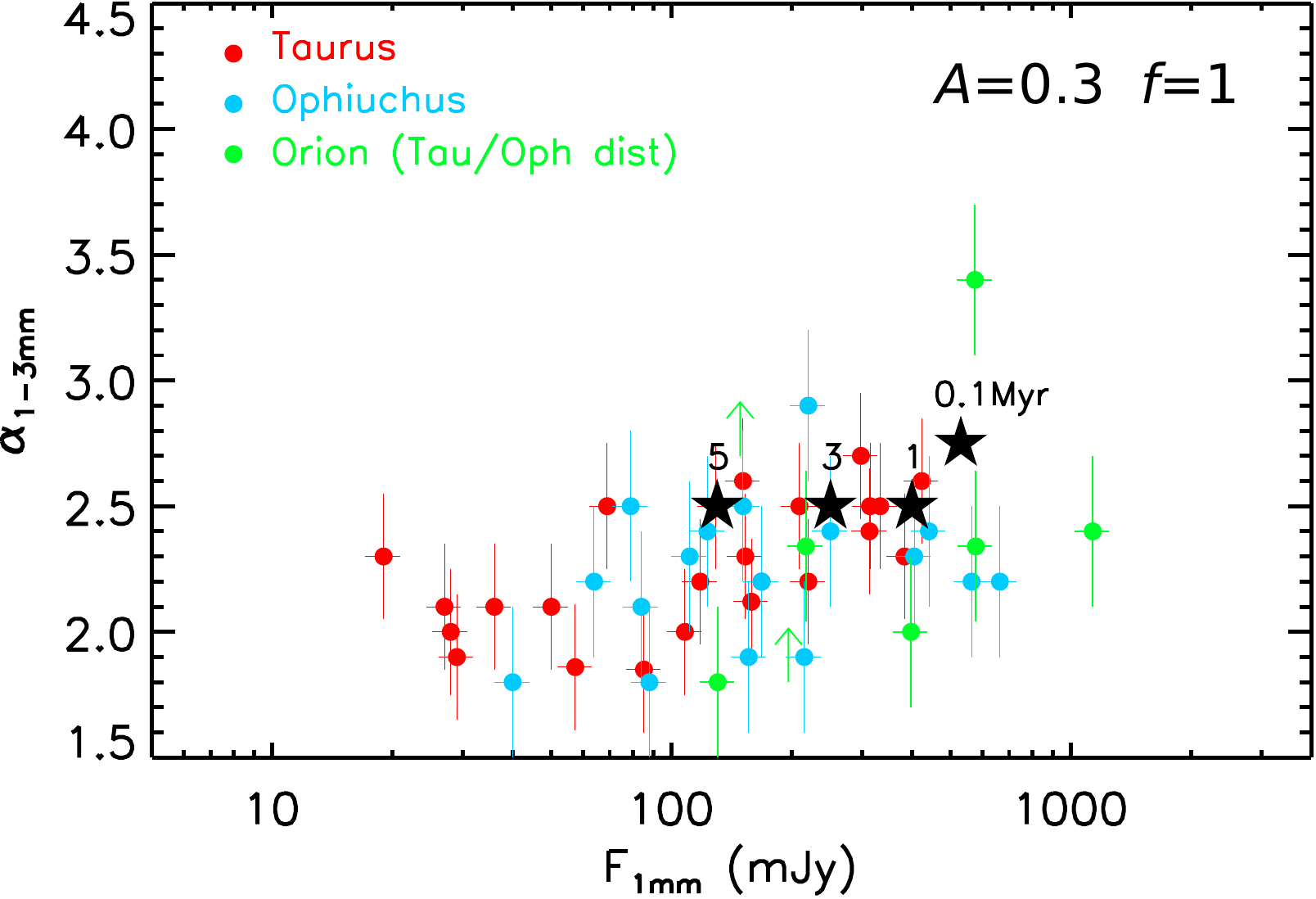}
   \caption{Comparison of the observed fluxes at mm-wavelengths of young disks in Taurus (red dots; from  \cite{2010A&A...512A..15R} and Ricci (priv comm)), Ophiucus (blue dots; from \cite{2010A&A...521A..66R}), and Orion Nebula Cluster (green dots; from \cite{2011A&A...525A..81R}) star forming regions with the predictions of the disk models at different times of the disk evolution (star symbols). Disk ages are indicated by numbers, in Myr, above the star symbols. The predicted $\sim1$ mm-fluxes (x-axis) and spectral indices between $\sim$1 and 3mm (y-axis) are for the disk models presented in Sect. \ref{sec2} with perturbations characterized by $f=1$ and $A=0.1$ (top panel) or $A=0.3$ (bottom panel). The $\sim 1$mm-flux densities for the Orion disks have been scaled by a factor of (420pc/140pc)$^2$ to account for the different distances estimated for the Orion Nebula Cluster ($\sim 420$pc, \cite{2007A&A...474..515M}) and Taurus and Ophiucus star forming regions ($\sim 140$pc, \cite{1999A&A...352..574B}, \cite{2008hsf2.book..351W})}
   \label{Fig3}
\end{figure}

The opacity, in the millimeter wavelength regime, can be approximated by a power law \citep{1993Icar..106...20M}, which means that the flux can be approximated to $F_\nu\propto\nu^{\alpha_{\mathrm{mm}}}$, where $\alpha_{\mathrm{mm}}$ is known as the spectral index. The spectral index gives us information about the size distribution of the dust in the disk. Figure \ref{Fig3} shows the time-dependent predicted fluxes at $\sim 1$mm ($F_{\mathrm{1mm}}$) and spectral index between $\sim 1$ and $3$mm ($\alpha_{\mathrm{1-3mm}}$) for a disk model with $f=1$ and $A=0.1$ (top panel) or $A=0.3$ (bottom panel). In the same plot mm-data for young disks in Taurus \cite{2010A&A...512A..15R} and Ricci (priv comm), Ophiucus \citep{2010A&A...521A..66R} and Orion Nubula Cluster \citep{2011A&A...525A..81R} are also shown.

%%%%%%%%%%%%
%FIGURE 9
%%%%%%%%%%%%
\begin{figure}
   \centering
  \includegraphics[width=8.8cm]{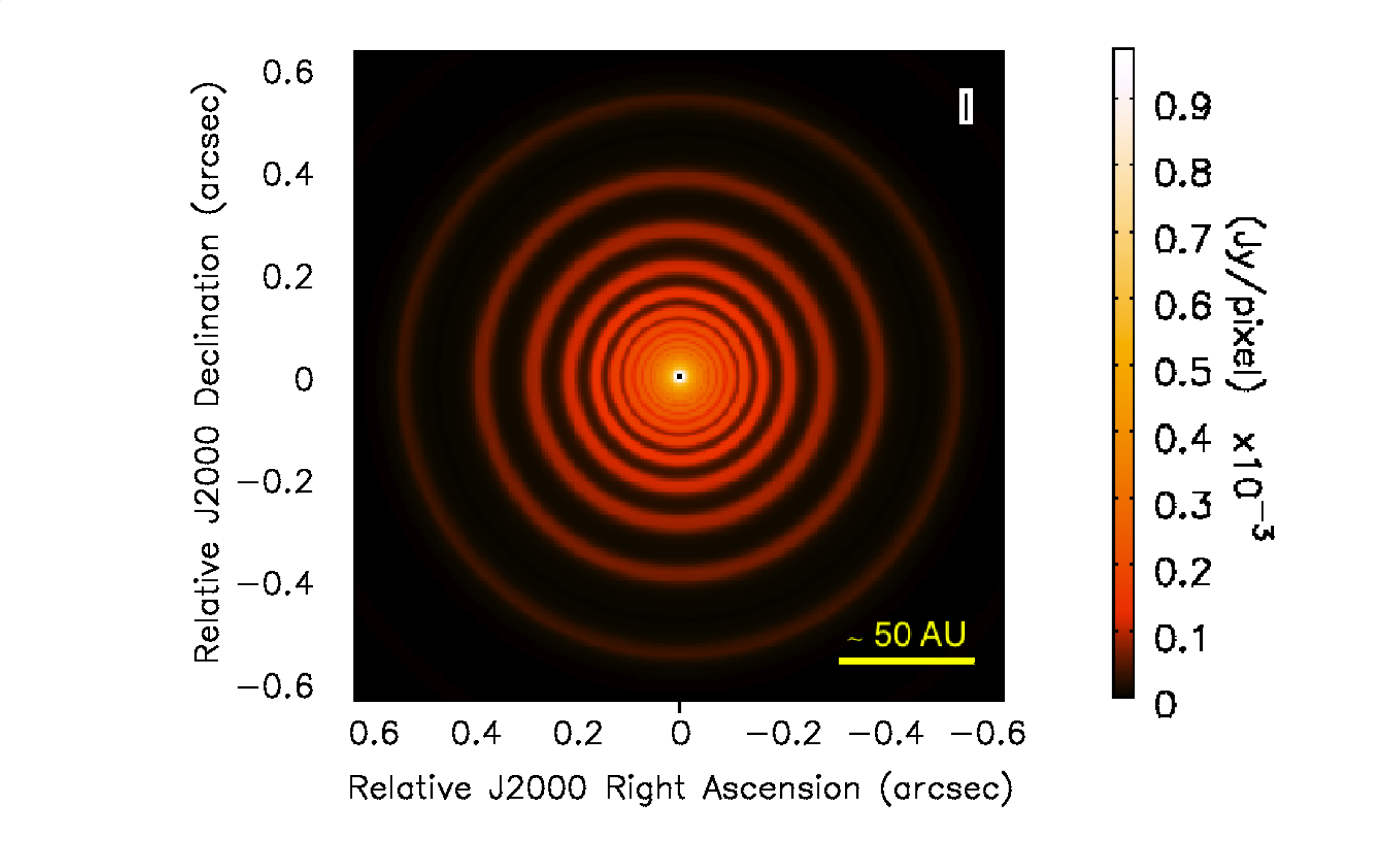}\\
  \includegraphics[width=8.8cm]{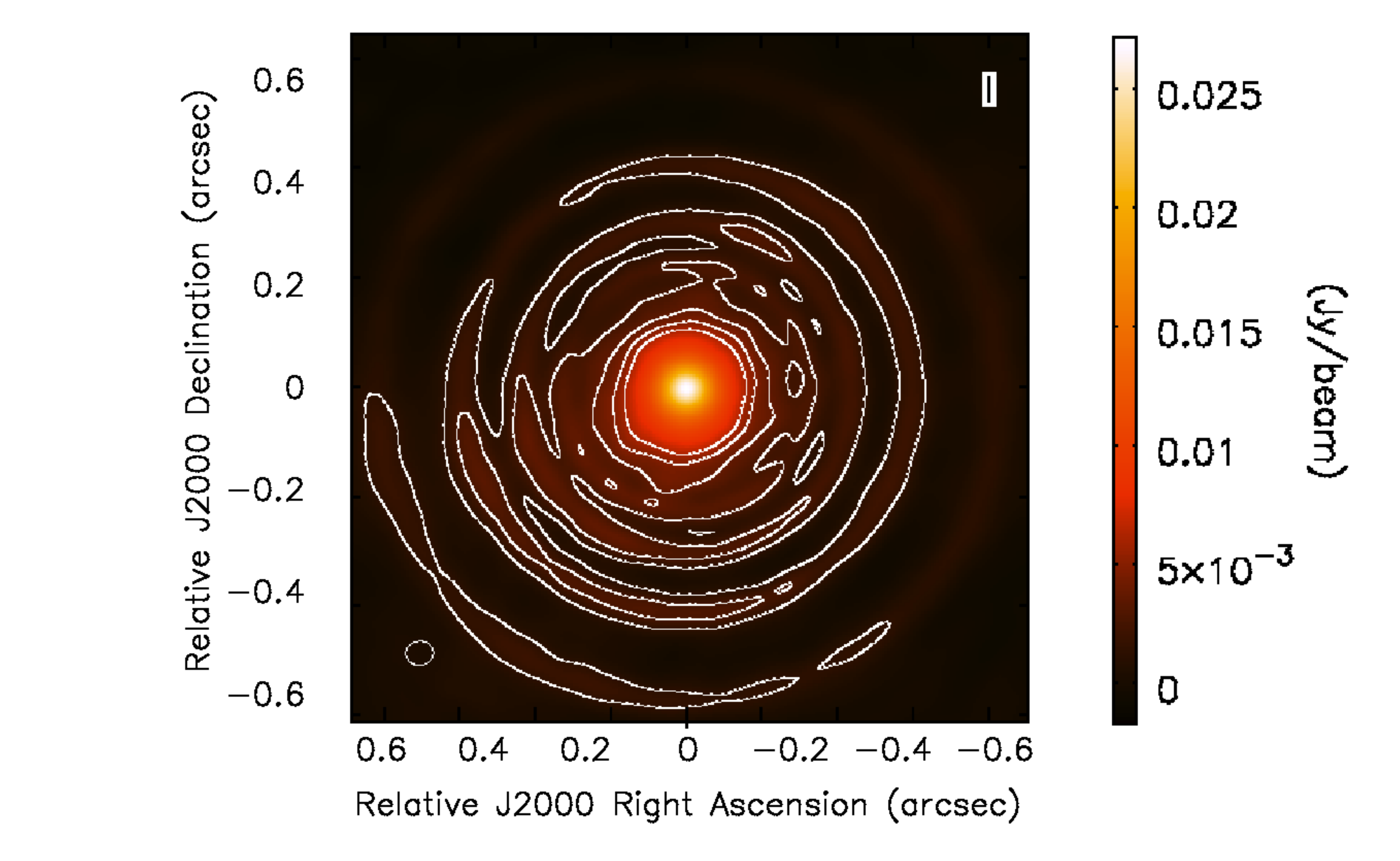}
   \caption{Disk image at 2 Myr and observing wavelength of 0.45 mm, the amplitude of the perturbation is $A=0.3$ and the factor $f=1$ for: disk model  with parameters of Table \ref{table1} (top), simulated image using  full configuration of ALMA (bottom)  with a maximum value of baseline of  around $3$km and  with an observing time of 4 hours. The contour plots are at $\{2, 4, 6, 8\}$ the corresponding rms value (see Table \ref{table:3}).}
              \label{Fig4}
 \end{figure}

As detailed in \cite{2010A&A...516L..14B} the $F_{\rm{1mm}}$ vs $\alpha_{\rm{1-3mm}}$ plot reflects some of the main properties of the dust population in the disk outer regions, which dominate the integrated flux at these long wavelengths. In particular the $1$mm-flux density is proportional to the total dust mass contained in the outer disk. The mm-spectral index is instead related to the sizes of grains: values lower than about 3 are due to emission from grains larger than about $\sim1 $mm, whereas values around 3.5 are due to smaller grains \citep{2007prpl.conf..767N}.

In the case of the disk model with $f=1$ and $A=0.1$, the time evolution of the predicted mm-fluxes clearly reflects the main features of the dust evolution depicted in top plot of Fig. \ref{Fig3}. Grains as large as a few millimeters are quickly formed in the disk outer regions ($R \gtrsim 50$~AU), and most of them are initially retained in those regions ($\lesssim 0.5$Myr). As a consequence, the mm-spectral index of the disk is slightly lower than 3 at these early stages. However, the radial drift of mm-sized pebbles becomes soon important and, as already described in Sect. \ref{sec3}, perturbations with a length-scale of $f=1$ and amplitude of $A=0.1$ are not efficient in retaining mm-sized particles in the outer disk. The 1 mm-flux density significantly decreases because of the loss of dust from the outer regions, especially the mm-sized grains which are efficient emitters at these wavelengths. Given that the spectral index is a proxy for the grain size, that is also affected by radial drift: its value increases with time because of the gradual loss of mm-sized pebbles in the outer disk. For this case, the under-predicted fluxes are not consistent with observational data for disk ages $\gtrsim 1$Myr, i.e. with the mean estimated ages of PMS stars in the Taurus, Ophiucus and Orion regions.  

%%%%%%%%%%%%
%FIGURE 10
%%%%%%%%%%%%
\begin{figure}
   \centering
  \includegraphics[width=8.8cm]{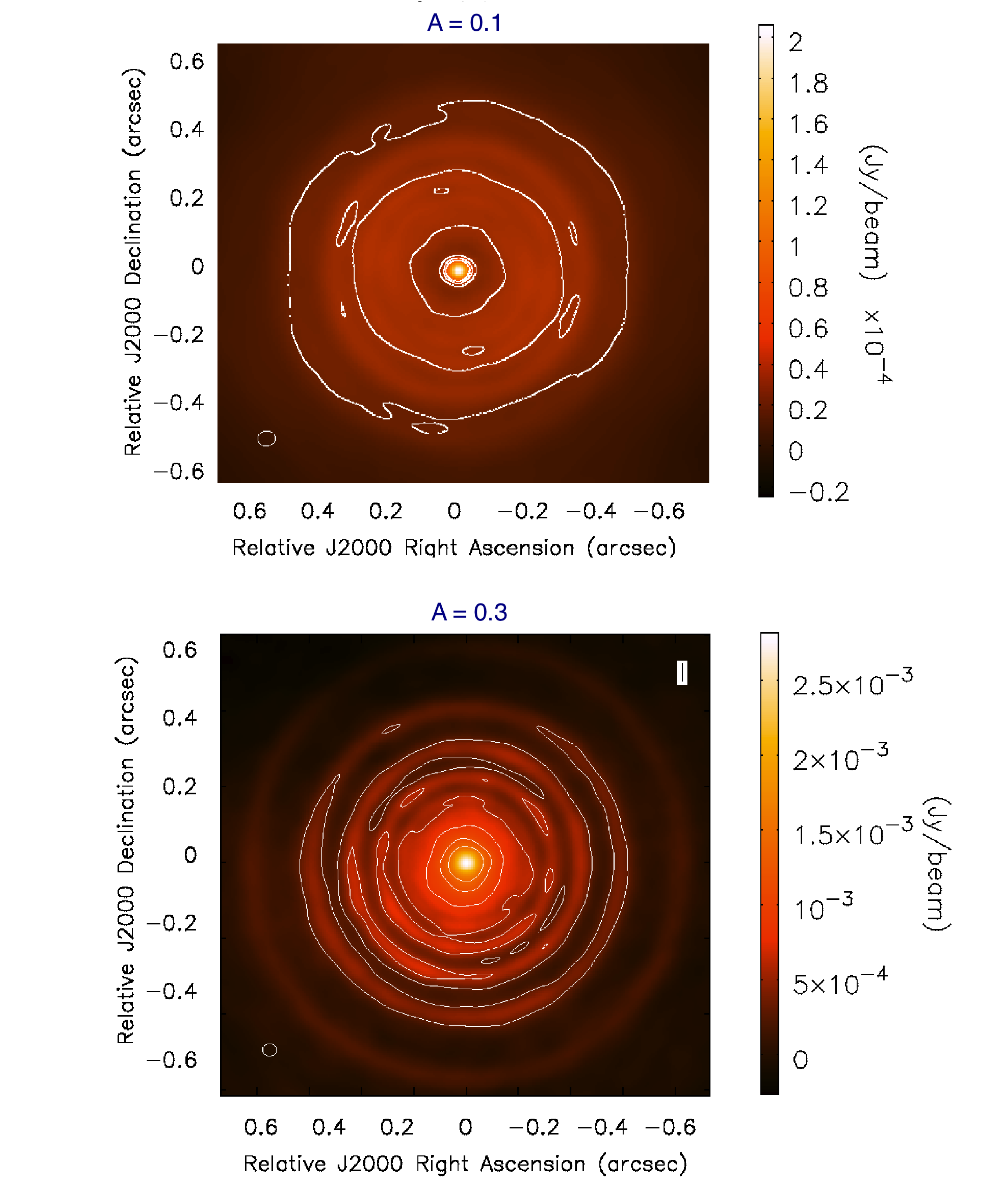}
   \caption{Comparison between the simulated images for an observing wavelength of $1$ mm and $2$  Myr of evolution, using full antenna configuration of ALMA for two different values of the amplitude of the perturbation: $A=0.1$ (top) and $A=0.3$ (bottom). The contour plots are at $\{2, 4, 6, 8\}$ the corresponding rms value (Table \ref{table:3} ).}
              \label{Fig6}
 \end{figure}

Interestingly, a disk with perturbations in the gas surface density with a larger amplitude of $A=0.3$ shows different results. In this case the trapping of particles in the pressure bumps is efficient enough to retain most of the large pebbles formed in the outer disk (see bottom plot of Fig.\ref{Fig3}). Since radial drift is much less efficient in this case, the predicted $1$mm-flux density is less affected than in the $A=0.1$ case and, more importantly, the spectral index levels off at a value of $\sim$2.5. This model provides a good match with the bulk of the mm-data for disk ages of a few Myr, as seen in bottom panel of Fig. \ref{Fig3}.

Note also that the match between the model presented here with $A=0.3$ is better than what has been obtained by \cite{2010A&A...516L..14B}. Contrary to the present work, they completely switched off radial drift, therefore they restricted particles to stay artificially in the disk outer regions. Specifically, for a disk with the same unperturbed disk structure presented here, they found a larger $ 1$mm-flux density than what we obtained in the $A=0.3$ perturbation case at few Myr. This indicates that in order to interpret the measured mm-fluxes of young disks both radial drift, and a physical mechanism acting in the disk to trap, although not completely, mm-sized particles in the outer disk are needed.

%%%%%%%%%%%%
%TABLE 2
%%%%%%%%%%%%
\begin{table*}
\caption{Atmospheric conditions, total flux and rms for the simulated observations at 140pc and at different observing wavelengths. The pwv value  takes into account the expected conditions for ALMA. The simulated images are using the full ALMA, but the antenna configuration is chosen in order to have the best  conciliation between resolution and sensitivity. }            
\label{table:3}      
\centering                         
\begin{tabular}{c c c c c}       
\hline\hline                
Amplitude   & Wavelength & Atmospheric conditions      		 &  Total Flux  			& rms\\ 
            	   &  (mm)             &pwv (mm)    $\quad\tau_0$              & (Jy)             			& (Jy)\\ 
\hline                
        		   &0.45	           &$0.5\qquad \qquad    0.60$				 &$6.9\times 10^{-1}$	&$7.5\times10^{-4}$ \\
A=0.3           &0.66	           &$1.0\qquad  \qquad    0.40$				 &$7.6\times 10^{-1}$	&$3.8\times10^{-4}$ \\
                      &1.00	           &$1.5\qquad  \qquad 0.20$				 &$2.2\times 10^{-1}$	&$1.4\times10^{-4}$ \\
		   &3.00	           &$2.3\qquad   \qquad 0.03$				 &$4.0\times 10^{-1}$	&$2.0\times10^{-5}$ \\
\hline
A=0.1           &1.00	           &$1.5\qquad  \qquad 0.20$				 &$1.6\times 10^{-2}$	&$1.3\times10^{-5}$ \\
\hline
\hline
\end{tabular}
\end{table*}

%%%%%%%%%%%%
%FIGURE 11
%%%%%%%%%%%%
\begin{figure*}
   \centering
  \includegraphics[width=18cm]{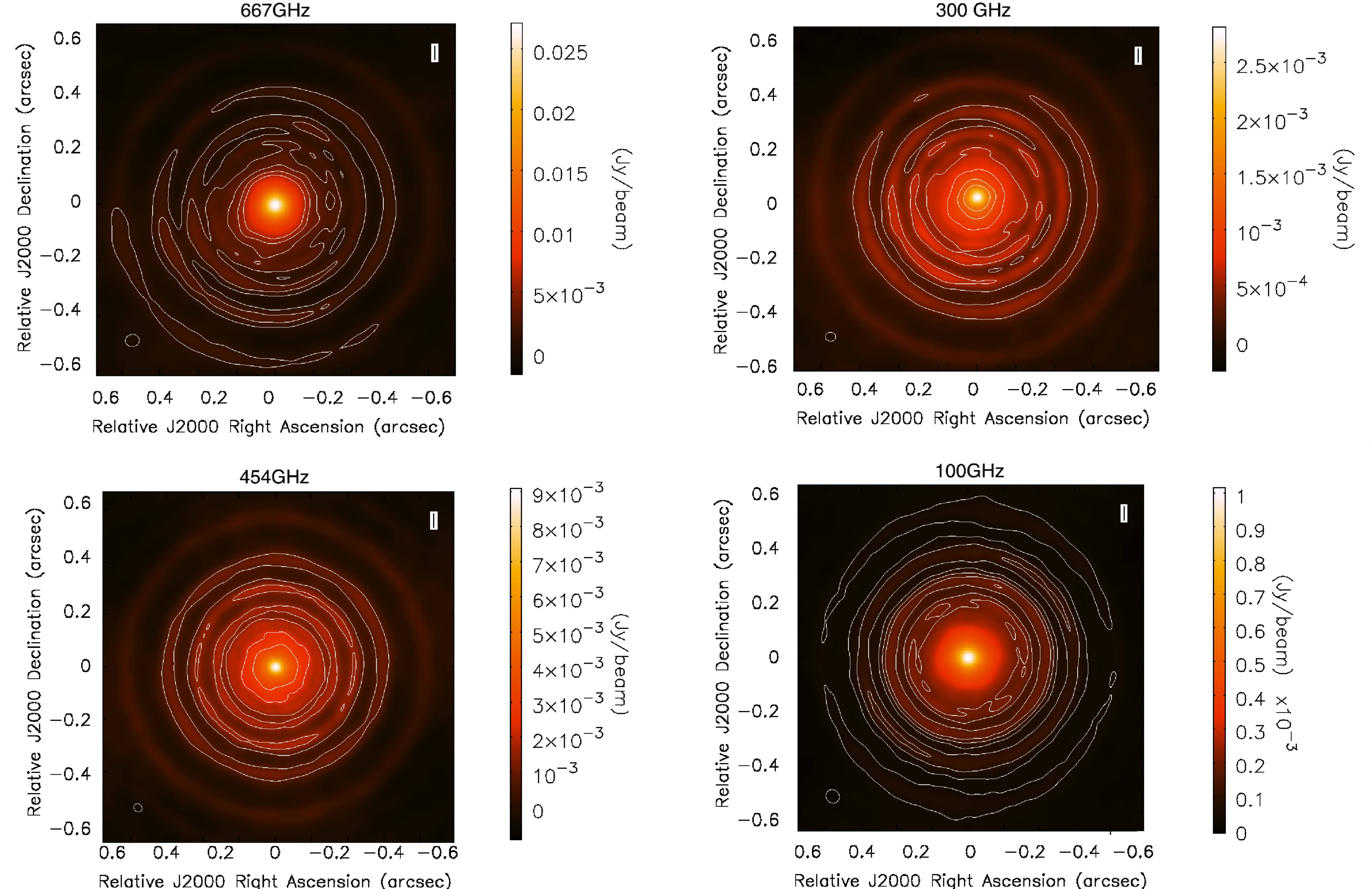}
   \caption{Disk simulated images with  parameters of Table \ref{table1}, $A=0.3$ and $f=1$ at 2 Myr of the disk evolution and for observing frequency of: $667$ GHz with a maximum baseline of  around 3 km (top left), $454$ GHz (bottom left) with a maximum baseline of around 4 km, $300$ GHz (top left)  with a max. baseline of around 7 km and $100$ GHz with a maximum baseline of 16 km (bottom right).   The  contour plots are at $\{2, 4, 6, 8\}$ the corresponding rms value (see Table \ref{table:3})}
              \label{compare}
 \end{figure*} 

%%%%%%%%%%%%
%FIGURE 12
%%%%%%%%%%%%
 \begin{figure}
   \centering
  \includegraphics[width=8.8cm]{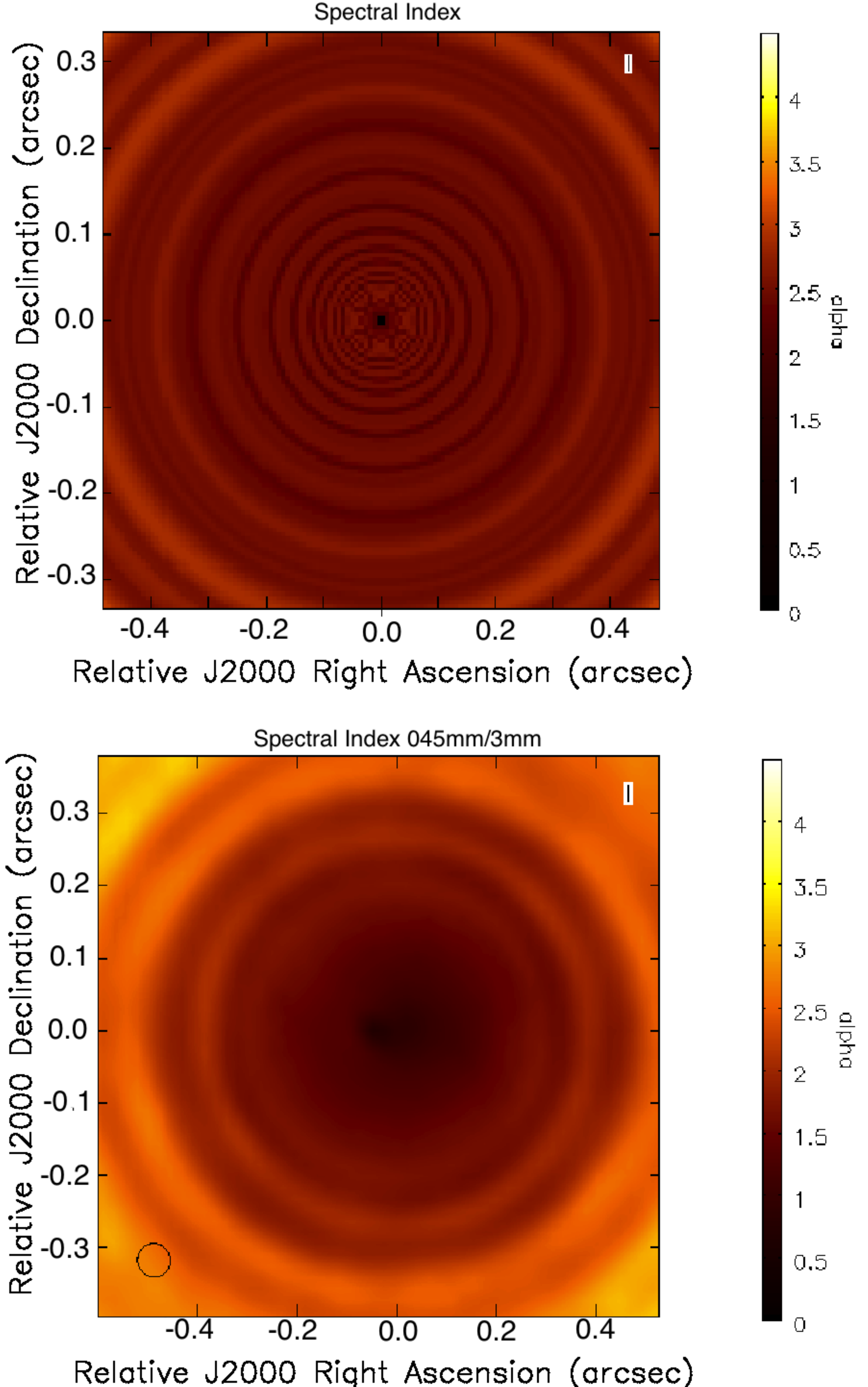}
   \caption{Spectral index $\alpha_{\mathrm{1-3mm}}$ of the model data (top plot) and the spectral index taking two simulated images at $0.45$mm and $3.0$mm, with a time observation of four hours and using the full configuration (max. baseline of 12 km)}
              \label{Fig5}
    \end{figure}
   
%%%%%%%%%%%%%
\subsection{Future observations with ALMA} \label{subsec3.3}
%%%%%%%%%%%%%
 
The Atacama Large Millimeter/sub-millimeter  Array (ALMA) will provide an increase in sensitivity and resolution  to observe in more detail the structure and evolution of protoplanetary disks. With a minimum beam diameter of $\sim 5$ mas at $900$GHz, ALMA will offer a resolution down to $2$ AU for disks observed in Orion and sub-AU for disk in Taurus-Auriga \citep{2010MNRAS.407..181C}. Using the Common Astronomy Software Applications (CASA) ALMA simulator (version 3.2.0), we run simulations to produce realistic ALMA observations of our model using  ALMA array of $50$ antennas $12$m-each.

The selection of observing mode to obtain the images, was chosen to have simultaneously the most favorable values for the resolution and sensitivity that should be available with ALMA. The spatial resolution  depends on the observing frequency and the maximum baseline of the array. We do not take the largest array because for very large baselines, the sensitivity could be not enough for the  regions that we will need to observe.  Therefore, we used different antenna arrays depending on the observing frequency to get the best possible resolution with enough sensitivity. The sensitivity depends on the  number of antennas,  the bandwidth (which is  taken as $\Delta \nu=8$GHz for continuum observations) and the total observing time that is fixed to four hours for each simulation.  The sensitivity also depends on the atmospheric conditions.  ALMA is located in Llano de Chajnantor Observatory, where  the precipitable water vapor (pwv) varies  between $0.5$ mm and $2.0$ mm depending on the observable frequency. For the simulations, we assume that the value of the pwv varies with frequency (see Table \ref{table:3}). The synthetic images are fully consistent with the opacity dust distribution discussed in Sec. \ref{subsec3.2}.

Figure \ref{Fig4} shows a comparison between the model image and a simulated ALMA image using the full configuration of ALMA with a maximum baseline of around $3$ km and an observation total time of four hours.  This image is for observing wavelength of $0.45$mm (Band 9 of ALMA $620-750$ GHz).   It is important to note that the simulated images take into account the atmospheric conditions and the expected receiver noise based on technical information of the antennas, but the residual noise after data calibrations and its uncertainties are not considered. We can note  (see Fig. \ref{Fig4}-bottom plot) that  with one of the full configurations  (max. baseline  $\sim 3$ km), it is possible to distinguish some  ring structures due to the fact that  the dust has drifted considerably into the rings compared to the gas.

In Fig. \ref{Fig6} we note again the importance of having a high value of the amplitude. If the gas surface density  of the disk is $A=0.3$, then the effects will be observable with ALMA. Both images of the figure have been compute with the complete antenna configuration of ALMA  for an observing wavelength of $1$ mm and $2$  Myr of the disk evolution. We can see that for $A=0.1$ it is not possible to detect any structures around the star even considering a perfect data calibration. It is important to note that due to the trapping of dust particles at peaks of the pressure bumps, the contrasts between rings in the simulated images is very clear, around $\sim 20-25\%$, while the contrast for the gas is almost  unrecognized.
 
Figure \ref{compare} compares the simulated images at different observing wavelength  using different antenna configurations of ALMA.  The antenna configuration is chosen by CASA depending on the expected resolution. The best image is obtained at $100$ GHz and a maximum baseline of 16 km (most extended ALMA configuration), where it is possible to detect clearly the most external ring structure and some internal ring structures. Nevertheless, with more compact configurations at different frequencies is still possible to detect some structures from the presence of the pressure bumps which allow  the formation of mm-sized particles. However, it is important to take into account that  the simulated images of Figs.  \ref{Fig4}, \ref{Fig6}  \ref{compare} and \ref{Fig5} are considering a perfect data calibration after  observations, for long baselines and high frequencies the calibration effects become more important.
 
Taking the ratio of the images at two different wavelengths, we will have the values of the spectral index $\alpha_{1-3mm}$, which indicates the location of mm-sized grains.  With the full configuration of ALMA and a maximum value of the baseline of $12$ km for both observing frequencies, some regions with large particles are  distinguished which are the regions with low spectral index $\alpha_{\mathrm{1-3mm}}\lesssim 3$, as it was explained in Sec. \ref{subsec3.2}. In Fig. \ref{Fig5} is the spectral index of the model data (top plot) and the spectral index taking two simulated images at $0.45$mm and $3.0$mm, with a total  observing time of four hours.

%%%%%%%%%%%%
%FIGURE 13
%%%%%%%%%%%%
 \begin{figure}
   \centering
  \includegraphics[width=8.8cm]{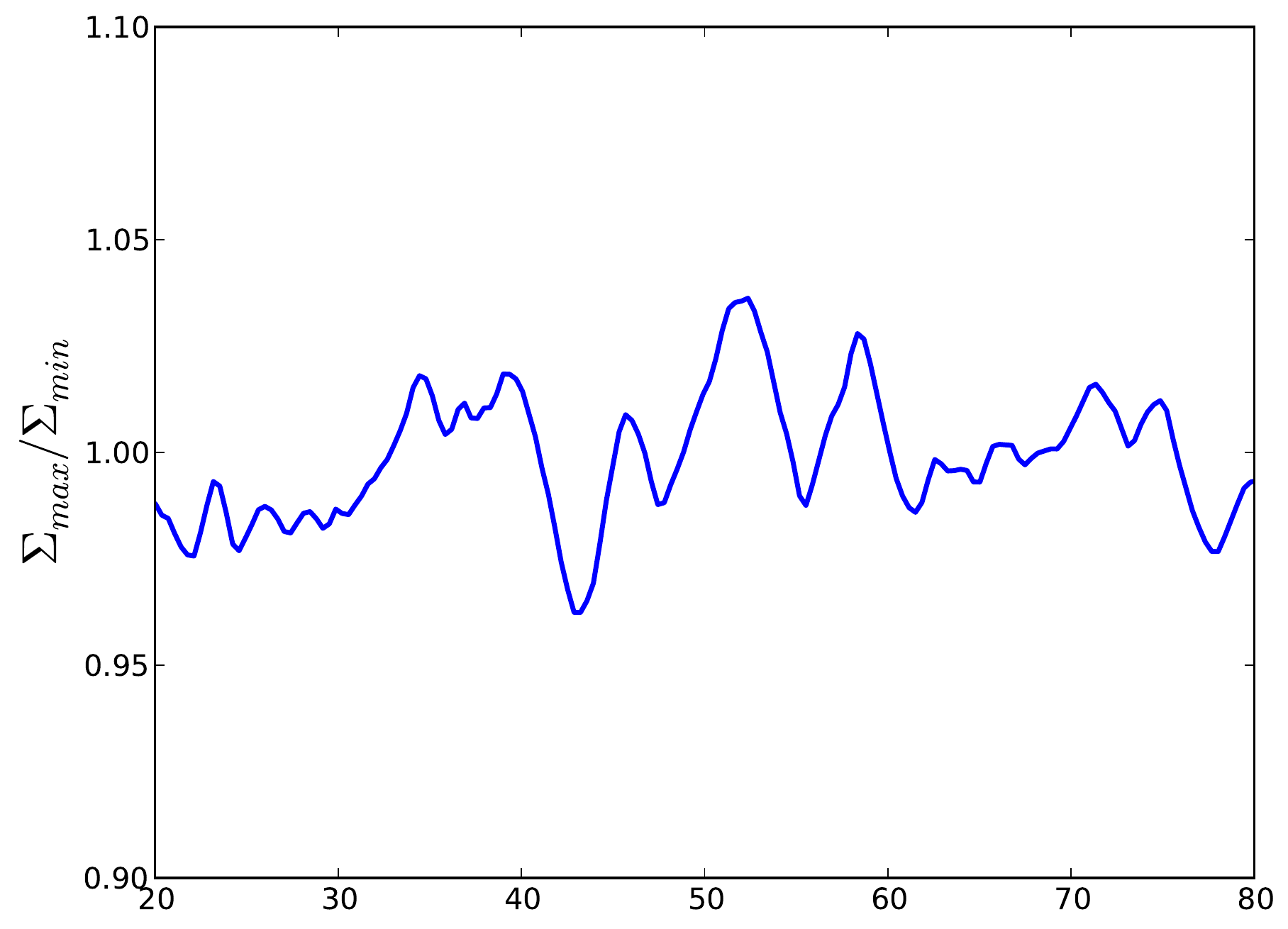}\\
  \includegraphics[width=8.8cm]{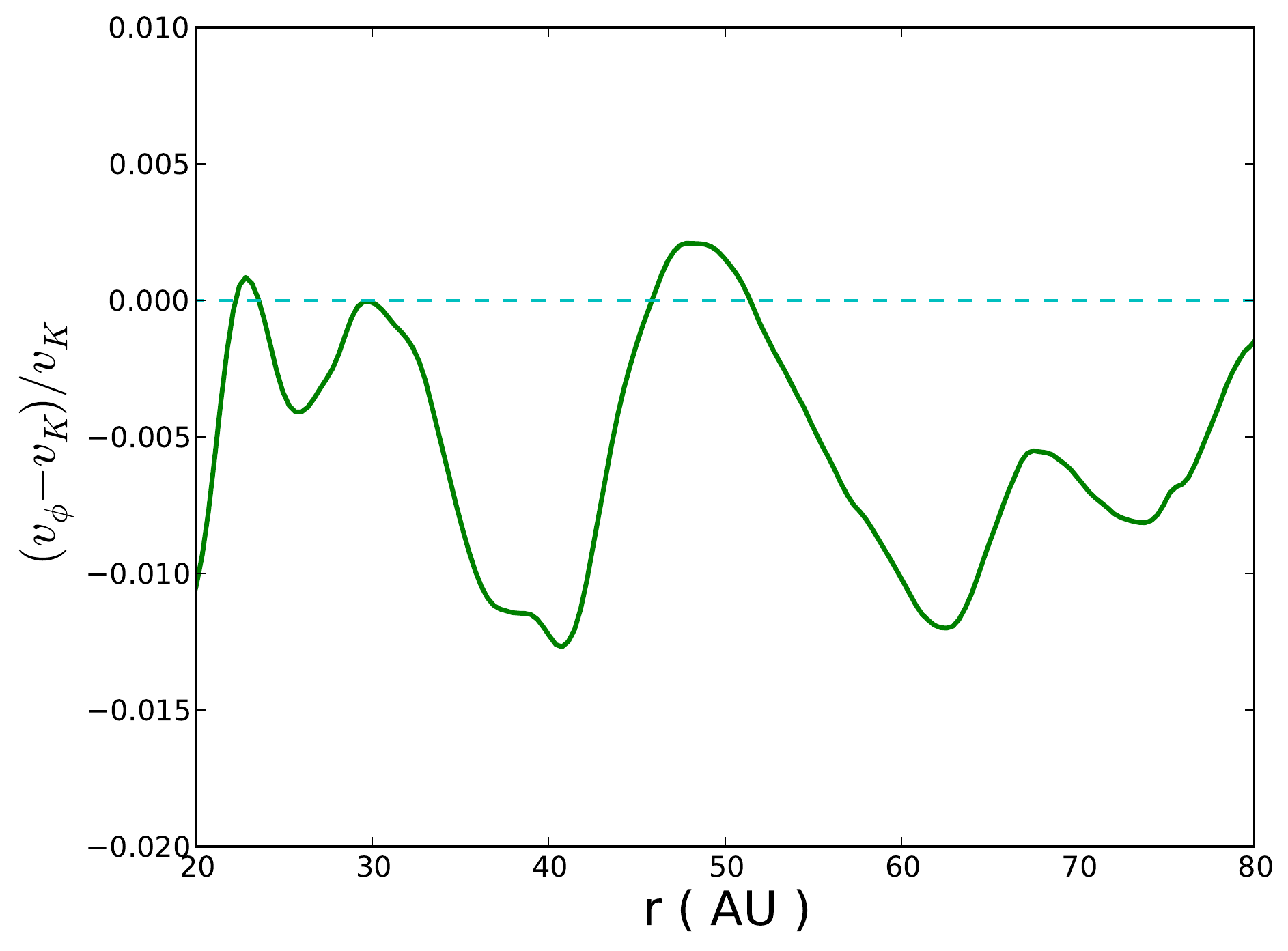}
    \caption{Top plot: Ratio between the surface density at two different azimuthal angles of of the disk from zonal flows simulation by  \citep{ 2011ApJ...736...85U}. The azimuthal angles are chosen such that for a specific radius, the amplitude of the pressure bump has a maximum $\Sigma_{\mathrm{max}}$ and a minimum $\Sigma_{\mathrm{min}}$. Bottom plot: Azimuthal velocity with respect to Keplerian velocity for the azimuthal and time-averaged surface density of the midplane.}
              \label{ratio_zonal}
 \end{figure}
 
%%%%%%%%%%%%%
\section{Approach to zonal flows predictions} \label{sec4}
%%%%%%%%%%%%%

In this paper we have so far assumed ad-hoc models of pressure bumps. But which processes may cause such long lived bumps in protoplanetary disks? In this section we study zonal flows as a possible explanation of the origin of long lived pressure bumps.

One possible cause of pressure bumps originates from MRI turbulence. \cite{1995ApJ...440..742H}  and \cite{1995ApJ...446..741B} were the first attempts to simulate nonlinear evolution of MRI in accretions disk, taking a box as representation of a small part of the disk. More recent simulations have been done taking higher resolution \citep[see e.g.][]{2009ApJ...697.1269J} and global setup \citep{2011ApJ...735..122F}.

%%%%%%%%%%%%%
%FIGURE 14
%%%%%%%%%%%%
 \begin{figure*}
   \centering
  	\includegraphics[width=18.0cm]{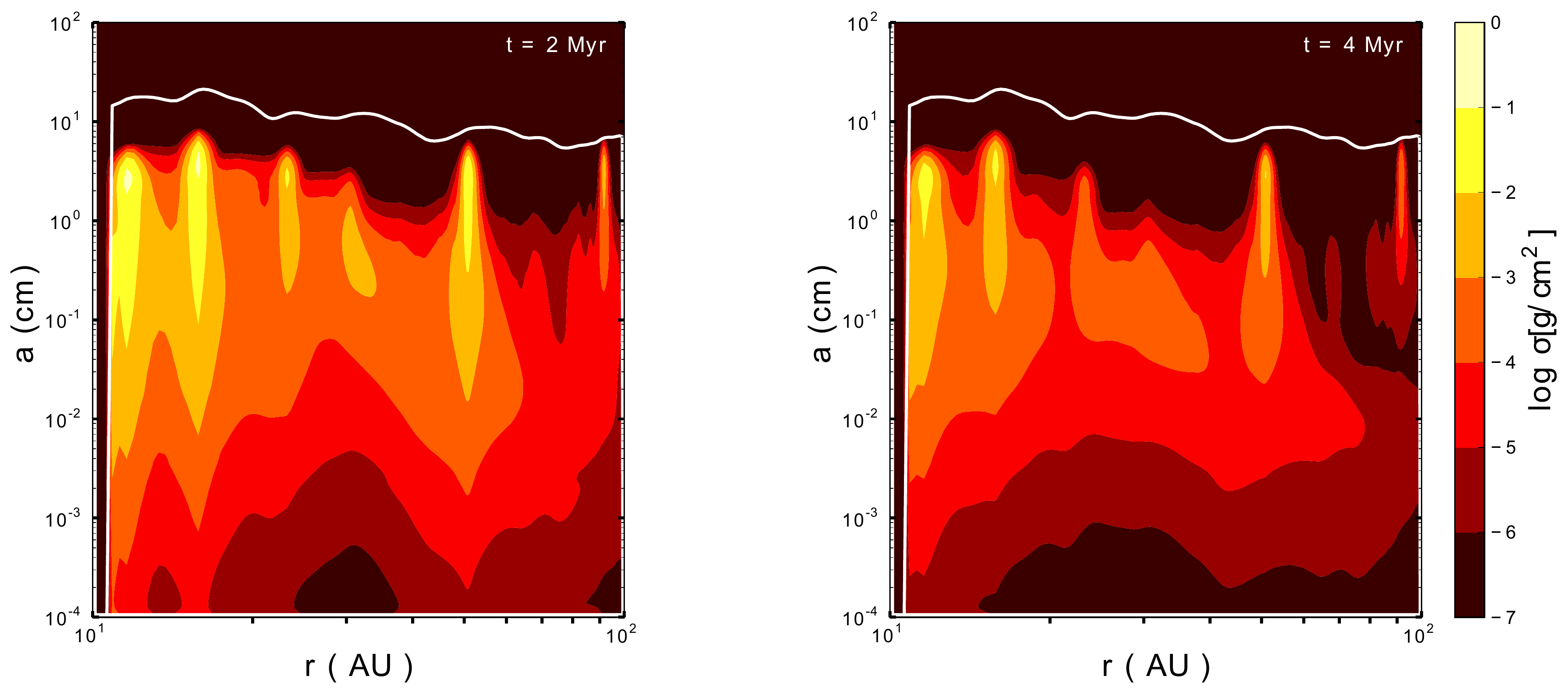}
    \caption{Vertically integrated dust density distribution after 2 Myr  and 4 Myr of dust evolution; taking the azimuthal and time-average profile of the gas surface density  in the midplane from the MHD simulations \cite[see][figure~3]{2011ApJ...736...85U} without a planet. The solid white line shows the particle size corresponding to a Stokes number of unity, which has the same shape of the density profile.}
              \label{zonal}
 \end{figure*}

In magnetorotational instability,  ``zonal flows''  are excited as a  result of the energy transportation from the MRI unstable medium scales, to the largest scales,  causing an inverse cascade of magnetic  energy, and creating a large-scale variation in the Maxwell stress \citep{2009ApJ...697.1269J}.  

Different three-dimensional MHD simulations show that with the presence of zonal flows, pressure bumps can appear if  there are drops in magnetic pressure through the disk \cite{2009ApJ...697.1269J}, \cite{2010A&A...515A..70D} and \cite{2011ApJ...736...85U}. Nevertheless, in recent simulations with higher resolution by  \cite{2011ApJ...735..122F}, pressure bumps are not formed. There is still not  a final answer about how and why these pressure bumps can or cannot be created via zonal flows.  

Another alternative for the origin of pressure bumps due to MRI is the change of the degree of ionization. The disk becomes MRI active if the degree of ionization is sufficient, for the magnetic field to be well couple to the gas. Variable degrees of ionization in the disk could cause local changes in the magnetic stress, which could induce structures in the density and pressure. 

The question we now wish to answer is: are the pressure bumps caused by zonal flows of MRI-turbulence strong enough to trap the dust like in the models of Sect. \ref{sec3}? To find this out, we take the three-dimensional MHD simulations of MRI-turbulent protoplanetary disks by \cite{2011ApJ...736...85U}. These models  have a resolution of $(N_r, N_{\theta}, N_{\phi})=(256,128, 256)$. It is important to notice that the MRI dynamical timescale is around one orbital period (at one AU) while the dust growth timescale is more than 100 orbital periods. Currently, it is not feasible to study the dust growth process self-consistently in the full time-dependent 3-D model. It would require running the MHD model tens to hundreds of times longer than what is currently affordable. Hence, the strategy is to find first a quasi-steady state of the gas surface density from MRI evolution, in which structures in the pressure survive the entire simulation (around 1000 inner disk orbits). Afterwards, to do the coagulation/fragmentation simulation of the dust in $1$D taking the gas surface density for a specific azimuthal angle in the midplane from the results from MHD simulations.

Top plot of Fig. \ref{ratio_zonal} shows the ratio between the time averaged surface density at two different azimuthal angles where the amplitude of a pressure bump is maximum and minimum for a specific radius. We can notice, that the variations on the azimuthal angle are very uniform, around $\sim 5 \%$. This is the reason why we work with the azimuthally averaged density.

For our simulations, we assume the pressure structure survives and is stable on dust growth and evolution timescales. The lifetime of these structures as determined by global disk simulations is still uncertain, but it has been observed to be on the order of 10-100 local orbits (at the radial position of the bump) \citep{2009ApJ...697.1269J, 2011ApJ...736...85U}. It is still an open question whether this behavior can be directly re-scaled to apply to the outer parts of the disk and in any case, the structures should be eventually diffused on turbulent diffusion timescales. In the future, the continuous generation and evolution of these structures should be implemented alongside the dust evolution. However, for lack of a better model of this time-dependence at this stage, we assume these structures to be static.

Since these MHD simulations use a radial domain where $r \in [1; 10]$, we rescale this grid logarithmically, so the gas surface density is taken from $10$AU to $100$AU, and also the surface density has been scaled such that the total disk mass is $0.05 M_\odot$, see Fig. \ref{comparacion}-left plot (solid-line). Comparing the gas surface density  obtained from MHD simulations with the assumed perturbed density $\Sigma'$ (Eq. \ref{eq1}), we see that the amplitude of the surface density perturbation from zonal flows is around $25\%$ and comparable with the amplitude of $30\%$ of $\Sigma '$. The widths of the bumps from \cite{2011ApJ...736...85U} are not uniform, but our assumption of $f=1$ fit well with some of those bumps.

The bottom plot of Fig. \ref{ratio_zonal} shows the azimuthal velocity with respect to Keplerian velocity for the azimuthal and time-averaged surface density of the midplane. We can notice that the azimuthal velocity exceeds the Keplerian velocity for some regions of the disk. This implies that for those regions, the presence of zonal flows allow to have a positive pressure gradient leaving dust particles to move outwards. Therefore, the peaks of the pressure bumps created by zonal flows may be regions where dust particles can reach millimeter sizes.

Figure \ref{zonal} shows the vertically integrated dust density distribution after 2 Myr and 4 Myr of dust evolution (making use of the model by \cite{2010A&A...513A..79B}). We can notice that at that time of evolution the pressure bumps caused by zonal flows are able to retain mm and cm sized particles in the outer regions of the disk. This is the reason why a high mass disk is considered in this case, in order to reach large grains. Around $50$-$60$AU there is clearly a region with high vertically integrated dust density distribution for mm and cm-sized particles. It is important to notice that the peak around $~100$AU is a result of the boundary condition.

%%%%%%%%%%%%%
%% CONCLUSIONS 
%%%%%%%%%%%%%

%%%%%%%%%%%%%
\section{Conclusions} \label{sec5}
%%%%%%%%%%%%%

Theoretical models of dust evolution in protoplanetary disks show that the growth   from sub-micron sized particles to larger objects is prevented basically by two phenomena: radial drift and fragmentation. Nevertheless, infrared and radio observations show that millimeter sized particles can survive under those circumstances in the outer regions of disks. Therefore, various theoretical efforts have been focused on explaining the survival of those bodies. 

Taking into account strong inhomogeneities expected to be in the gas density profile e.g. zonal flows,  and using the coagulation/fragmentation and disk-structure models by  \cite{2010A&A...513A..79B}, we have investigated how the presence of pressure bumps can cause the reduction of radial drift, allowing the existence of millimeter sized grains in agreement with observations. In this work, we assumed a sinusoidal function for the gas surface density  to simulate pressure bumps. The amplitude and wavelength disturbances are chosen considering the necessary conditions to have outward angular momentum transport in an $\alpha$-turbulent type disk, outward radial drift of dust and reasonable values compare to predictions from the recent work of zonal flows \citep{2011ApJ...736...85U}.  

The results presented here suggest that the presence of pressure bumps with a width of the order of the disk scale-height and an amplitude of $30\%$ of the gas surface density of the disk, provide the necessary physical conditions for the survival of larger grains in a disk with properties summarized in Table ~\ref{table1}.  Comparisons between the observed fluxes of the Taurus, Ophiucus and Orion Nebula Cluster star forming regions with the results of the models ratify that the effect of the radial drift is reduced allowing particles to grow. Figure \ref{Fig3}  shows how models with these kind of disturbances  reproduce much better mm-observations than models with full or without radial drift.

In addition, we presented a comparison between the bumpy density profile assumed in this work and 3D MHD models of zonal flows that can cause long lived bumps in protoplanetary disks. We showed that the pressure bumps cause by zonal flows of  \citep{2011ApJ...736...85U} are in agreement with the amplitudes and wavelengths used in this work. Therefore, taking those bumps, the survival of dust particles is possible in the outer regions after some Myr. 

The simulated images using CASA ALMA simulator (version 3.2.0) show that, with different antenna configuration of the final ALMA stage, the ring structures, due to the presence of the pressure bumps, should be detectable. Future ALMA observations will have an important impact for understanding the first stages of planet formation and it will  be very important to investigate if the  grain growth and retetion can be explained with the presence of these kind of inhomogeneities in the gas density profile. 

\begin{acknowledgements}
We acknowledge Francesco Trotta for his help with the code that we used in this work to derive the mm-fluxes. We would like to thank the referee, Wladimir Lyra, for his useful suggestions. This work was supported in part through the 3rd funding line of German Excellence Initiative.   T. Birnstiel  acknowledges ESO Office for Science which provided funding for the visits in Garching. L. Ricci acknowledges the PhD fellowship of the International Max-Planck-Research School.  A. L. Uribe acknowledges the CPU time for running the simulations in the Bluegene/P supercomputer and the THEO cluster at the Rechenzentrum Garching (RZG) of the Max Planck Society. Finally,  L. Testi acknowledges ASI contract to the INAF-Osservatorio Astrofisico Di Arcetri.  
\end{acknowledgements}

%%%%%%%%%%%%%
%% REFERENCES
%%%%%%%%%%%%%

\bibliographystyle{aa}  
\bibliography{Pinilla.bib}

%%%% END%%%%%%%

\end{document}